\documentclass[npg]{copernicus}
\usepackage{caption}
\usepackage{subcaption}

\begin{document}

\linenumbers

\title{Interaction of a monopole vortex with an isolated topographic feature in a three-layer geophysical flow}

\author[1]{Evgeny A. Ryzhov}
\author[1,2]{Konstantin V. Koshel}

\affil[1]{V.I.Il`ichev Pacific Oceanological Institute, 43, Baltiyskaya Street, Vladivostok, 690041, Russia}
\affil[2]{Far Eastern Federal University, 8, Sukhanova Street, Vladivostok, 690950, Russia}


\runningtitle{Monopole-topography interaction}

\runningauthor{E. A. Ryzhov and K. V. Koshel}

\correspondence{E. A. Ryzhov\\ (ryzhovea@poi.dvo.ru)}

\received{}
\pubdiscuss{} 
\revised{}
\accepted{}
\published{}


\firstpage{1}

\maketitle  

\begin{abstract}
In the frame of a three-layer quasi-geostrophic analytical model of a $f$-plane geophysical flow, Lagrangian advection being induced by the interaction of a monopole vortex with an isolated topographic feature is addressed. Two different cases when the monopole locates either within the upper or the middle layer are of our interest. In the bottom layer, there is a delta function topographic feature, which generates a closed recirculation region in its vicinity due to the background flow. This recirculation region extends to the middle and upper layers, and it plays the role of a topographic vortex. The interaction between the monopole and the topographic vortex causes complex, including chaotic, advection of fluid particles. We show that the model's parameters, namely, the monopole and topographic vortices' strengths and initial positions, the layers' depths and densities are responsible for the diverse advection patterns. While the patterns are rather complicated, however, one can single out two major processes, which mostly govern fluid particle advection. The first one is the variation in time of the system's phase space structure, so that within the closed region of the topographic vortex, there appear periodically unclosed particle pathways by which the particles leave the topographic vortex. The second one is chaotic advection that arises from the nonstationarity of the monopole-topography interaction.
\end{abstract}


\introduction  
Generally speaking, topographic vortices are coherent vortical structures appearing as closed recirculation regions over bottom features in the ocean and atmosphere. Topographic vortices play a fundamental role in mass, salinity and temperature advection in the ocean. Moreover, topographic vortices are known to influence the dynamics of different coherent structures, such as unrestrictedly moving vortices \citep[e.g.,][]{vanGeffen_Davies_1999, Dewar_2002, An_McDonald_2005, Candon_Marshall_2012, ZavalaSanson_et_al_2012}. Such topographic vortices greatly vary in time and size scales \citep{Baines_Smith_1993, Baines_1993}. In this paper, however, we are only interested in meso- and synoptic scale topographic vortices due to these scales are generally believed to be prevailing in the ocean  \citep{Chelton_et_al_2011}.

The present paper deals with Lagrangian regular and irregular (chaotic) advection being generated by a vortex monopole interacting with a topographic vortex. The topographic vortex under investigation is generated by a regular three-layer $f$-plane background flow \citep[e.g.,][]{Pedlosky_1987, Kozlov_1995} with a delta function bottom irregularity within the lower layer \citep[e.g.,][]{Sokolovskiy_et_al_1998, Izrailsky_et_al_2004, Kozlov_et_al_2005}. Then we embed a monopole singular vortex \citep[e.g.,][]{Gryanik_1983, Gryanik_Tevs_1989, Gryanik_et_al_2000, Carton_2001, Reznik_2010, Reznik_Kizner_2010} either within the upper or middle layer. So, these singularities move like passive tracers along regular background flow stream-lines \citep{Reznik_Kizner_2007a,Reznik_Kizner_2007b, Reznik_Kizner_2010}, although, generating a complex either periodically (for a time-independent background flow) or quasi-periodically (for a periodically time-dependent background flow) velocity field in the vicinity of themselves. Our main reason for employing such a three-layer model \citep[e.g.,][]{Sokolovskiy_1997, Ryzhov_Koshel_2011a} is to study Lagrangian advection being induced by the monopole-topography interaction when the monopole being located within either the upper or middle layers. Such monopole positioning can be considered as the simplest models for a surface eddy, and for an interthermocline lens  \citep[e.g.,][]{Carton_et_al_2002,Wang_Dewar_2003, Filyushkin_et_al_2011, Filyushkin_Sokolovskiy_2011}, respectively. However, we should emphasize the paper to deal only with surface Lagrangian advection in both cases. The middle-layer singular monopole appears on the upper layer as a regular vortex, so this configuration generates different fluid particle advection scenarios due to no singularities occur in the monopole-topography interaction velocity field.

We investigate two kinds of the monopole-topography interaction, the first one is an infinity-time interaction and the second one is a finite-time interaction. The infinite-time interaction means the monopole to move in closed regular trajectories about the topographic vortex's elliptic point for infinite time due to the constancy of the background flow. However, if the background flow depends periodically on time, the dynamics of the monopole becomes more complicated. So that now the monopole itself can be captured within the topographic vortex from the background flow or, on the contrary, be released from the topographic vortex into the background flow, and, consequently, be carried away to the infinity. 

Thus, the main aim of the present study is to investigate Lagrangian advection of fluid particles occurring due to the velocity field being generated by the infinite and short-term monopole-topography interactions.
\section{Model formulation}
The simplest way to study a quasi-two-dimensional layered geophysical flow is by exploiting the potential vorticity definition in each layer. For a background three-layer flow under the rigid lid approximation, these definitions read \citep{Pedlosky_1987}
\begin{eqnarray}
\label{eq1}
{q_1} &=& \Delta {\psi _1} + \frac{f}{{{H_1}}}{\zeta _1} + f,\; \; {q_2} = \Delta {\psi _2} + \frac{f}{{{H_2}}}\left( {{\zeta _2} - {\zeta _1}} \right) + f,\\ \nonumber
{q_3} &=& \Delta {\psi _3} + \frac{f}{{{H_3}}}\left( {h\left( {x,y} \right) - {\zeta _2}} \right) + f,
\end{eqnarray}
where $i=1,2,3$ corresponds to the upper, middle, and lower layer respectively; $q_i$ is the $i$-layer potential vorticity, $\Delta {\psi _i} = \frac{{\partial {v_i}}}{{\partial x}} - \frac{{\partial {u_i}}}{{\partial y}}$ is the two-dimensional relative vorticity with stream-function $\psi _i$ and two-dimensional velocity field $u_i,\, v_i$; $\zeta _1,\, \zeta _2$ are the interface heights between the upper and middle, and the middle and lower layers, respectively; $h(x,y)=\tau \delta (\bf{r}))$ is the Dirac delta function bottom irregularity with effective volume $\tau$; $H_i$ is the $i$-layer depth; $f$ is the constant Coriolis parameter. According to the pressure continuity condition, the interface heights can be written in the form \citep{Pedlosky_1987},
\begin{equation}
\label{eq2}
{\zeta _1} = \frac{{f\left( {{\psi _2} - {\psi _1}} \right){\rho _2}}}{{\left( {g\left( {{\rho _2} - {\rho _1}} \right)} \right)}},\;\;{\zeta _2} = \frac{{f\left( {{\psi _3} - {\psi _2}} \right){\rho _3}}}{{\left( {g\left( {{\rho _3} - {\rho _2}} \right)} \right)}},
\end{equation}
where $\rho _i$ is the $i$-layer fluid density; $g$ is the gravitational acceleration; $\Delta \rho _1 = \rho _2-\rho _1$, and $\Delta \rho _2 = \rho _3-\rho _2$ are the density jumps. 

Substituting (\ref{eq2}) into (\ref{eq1}), one can obtain the detailed potential vorticity, 
\begin{eqnarray}
\label{eq3}
{q_1} &=& \Delta {\psi _1} + {k_1}\left( {{\psi _2} - {\psi _1}} \right) + f,\\\nonumber
{q_2} &=& \Delta {\psi _2} + \left( {{k_{21}}{\psi _1} - {\psi _2}\left( {{k_{21}} + {k_{22}}} \right) + {k_{22}}{\psi _3}} \right) + f,\\\nonumber
{q_3} &=& \Delta {\psi _3} + {k_3}\left( {{\psi _2} - {\psi _3}} \right) + \frac{{f\tau }}{{{H_3}}}\delta \left( {\bf{r}} \right) + f,
\end{eqnarray}
where ${k_1} = \frac{{f^2{\rho _2}}}{{{H_1}g\Delta {\rho _1}}}$, ${k_3} = \frac{{f^2{\rho _3}}}{{{H_3}g\Delta {\rho _2}}}$, ${k_{21}} = \frac{{f^2{\rho _2}}}{{{H_2}g\Delta {\rho _1}}}$, ${k_{22}} = \frac{{f^2{\rho _3}}}{{{H_2}g\Delta {\rho _2}}}$. 

Since our study concerns only the cases of the upper- and middle-layer monopole propagation, we set the lower-layer potential vorticity to be always time-independent. However, either the upper or middle layer potential vorticity has one time-dependent singular value moving with the monopole's center. Hence, we have two sets of singular perturbations of the flow,
\begin{equation}
\label{eq4}
{q_m} = q_m^\ast + \frac{f}{{{H_m}}}{\mu _m}\delta \left( {\left| {{{\bf{r}}_i} - {\bf{r}}_m^\ast} \right|} \right),\;\;{q_n} = q_n^\ast,\;\;{q_3} = q_3^\ast,
\end{equation}
where $m,\;n =1,2$, $m \ne n$, $\mu _m$ is the monopole's strength and $\bf{r}_m ^\ast $ is the position of the monopole's singularity within the $m$-layer, $q _i ^\ast$ is the potential vorticity background value, and $\left| {{{\bf{r}}_i} - {\bf{r}}_m^*} \right| = \sqrt {{{\left( {{x_i} - x_m^*} \right)}^2} + {{\left( {{y_i} - y_m^*} \right)}^2}}$ with $x_i, \; y_i$ being Cartesian coordinates of a fluid particle within the $i$-layer.

Potential vorticity (\ref{eq1}) should satisfy the potential vorticity conservation law in each layer, 
\begin{equation}
\label{eq5}
{\partial _t}{q_i} + J\left( {{\psi _i},{q_i}} \right) = 0.
\end{equation}

To obtain explicit analytical relations for stream-functions $\psi_i$, one can split relations (\ref{eq3}) by making use of the following procedure \citep[e.g.,][]{Gryanik_Tevs_1989}. First, we rewrite (\ref{eq3}) in a matrix form, 
\begin{equation}
\label{eq6}
\Delta {\bf{\Psi}} + {\bf{A\Psi}} = {\bf{B}},
\end{equation}
where $\bf{\Psi} = \left( {\begin{array}{*{20}{c}}
{\psi_1}\\
{\psi_2}\\
{\psi_3}
\end{array}} \right)$, ${\bf{A}} = \left( {\begin{array}{*{20}{c}}
{ - {k_1}}&{{k_1}}&0\\
{{k_{21}}}&{ - \left( {{k_{21}} + {k_{22}}} \right)}&{{k_{22}}}\\
0&{{k_3}}&{ - {k_3}}
\end{array}} \right)$, and ${\bf{B}} =  - \left( {\begin{array}{*{20}{c}}
{f + {q_1}}\\
{f + {q_2}}\\
{f + {q_3}}
\end{array}} \right)$. Second, we diagonalize matrix $\bf{A}$ through a similarity transformation, ${\bf{A}} = {\bf{SJ}}{{\bf{S}}^{ - 1}}$. Matrix $\bf{S}$, whose columns are the eigenvectors of matrix $\bf{A}$, and diagonal matrix $\bf{J}$, whose main diagonal is consisted of the eigenvalues of matrix $\bf{A}$, have the form,
\begin{equation}
\label{eq7}
{\bf{S}} = \left( {\begin{array}{*{20}{c}}
1&{{\alpha _1}}&{{\beta _1}}\\
1&{{\alpha _2}}&{{\beta _2}}\\
1&1&1
\end{array}} \right),\;\;{\bf{J}} = \left( {\begin{array}{*{20}{c}}
0&0&0\\
0&{ - {k_3}\left( {{\alpha _2} - 1} \right)}&0\\
0&0&{ - {k_3}\left( {{\beta _2} - 1} \right)}
\end{array}} \right),
\end{equation}
where
\begin{eqnarray}
\label{eq8}
{\alpha _1} &=&  - {k_{22}}/{k_{21}} - {\alpha _2}/{k_{21}}\left( { - {k_{21}} - {k_{22}} + {k_3}\left( {{\alpha _2} - 1} \right)} \right),\\\nonumber
{\alpha _2} &=& \left( {{k_1} + {k_3} + {k_{21}} + {k_{22}} + {\lambda _0}} \right)/\left( {2{k_3}} \right),\\\nonumber
{\beta _1} &=&  - {k_{22}}/{k_{21}} - {\beta _2}/{k_{21}}\left( { - {k_{21}} - {k_{22}} + {k_3}\left( {{\beta _2} - 1} \right)} \right),\\\nonumber
{\beta _2} &=& \left( {{k_1} + {k_3} + {k_{21}} + {k_{22}} - {\lambda _0}} \right)/\left( {2{k_3}} \right),\\\nonumber
{\lambda _0} &=& \sqrt {{{\left( {{k_1} - {k_3} + {k_{21}} + {k_{22}}} \right)}^2} - 4\left( { - {k_1}{k_3} - {k_3}{k_{21}} + {k_1}{k_{22}}} \right)}.
\end{eqnarray}
Third, we introduce vector $\bf{\Phi}$ such that $\bf{\Psi}=\bf{S}\bf{\Phi}$, and obtain from (\ref{eq6}) new expression ${\Delta\bf{\Phi}} + {\bf{J\Phi }} = {{\bf{S}}^{ - 1}}{\bf{B}}$, which, taking into consideration relations ${\bf{S}^{ - 1}} = \frac{1}{\gamma }\left( {\begin{array}{*{20}{c}}
{{\alpha _2} - {\beta _2}}&{{\beta _1} - {\alpha _1}}&{{\alpha _1}{\beta _2} - {\alpha _2}{\beta _1}}\\
{{\beta _2} - 1}&{1 - {\beta _1}}&{{\beta _1} - {\beta _2}}\\
{1 - {\alpha _2}}&{{\alpha _1} - 1}&{{\alpha _2} - {\alpha _1}}
\end{array}} \right)$, and $\gamma  = {\alpha _2} - {\alpha _1} + {\beta _1} - {\beta _2} + {\alpha _1}{\beta _2} - {\alpha _2}{\beta _1}$, has the following detailed form, 
\begin{eqnarray}
\label{eq9}
\Delta {\Phi _1} &=&  - f - \frac{1}{\gamma }\left[ \left( {{\alpha _2} - {\beta _2}} \right){q_1} + \left( {{\beta _1} - {\alpha _1}} \right){q_2} + \right.\\ \nonumber
 && +\left. \left( {{\alpha _1}{\beta _2} - {\alpha _2}{\beta _1}} \right){q_3} \right],\\ \nonumber
\Delta {\Phi _2} &-& {k_3}\left( {{\alpha _2} - 1} \right){\Phi _2} =  - \frac{1}{\gamma }\left[ \left( {{\beta _2} - 1} \right){q_1} + \left( {1 - {\beta _1}} \right){q_2} + \right.\\ \nonumber
 && +\left. \left( {{\beta _1} - {\beta _2}} \right){q_3} \right],\\ \nonumber
\Delta {\Phi _3} &-& {k_3}\left( {{\beta _2} - 1} \right){\Phi _3} =  - \frac{1}{\gamma }\left[ \left( {1 - {\alpha _2}} \right){q_1} + \left( {{\alpha _1} - 1} \right){q_2} + \right.\\ \nonumber
 && +\left. \left( {{\alpha _2} - {\alpha _1}} \right){q_3} \right].
\end{eqnarray}
The last step is to obtain explicitly barotropic mode $\Phi _1$ and two baroclinic modes $\Phi _2$, $\Phi _3$. As it has been mentioned above, we are interested only in the singular perturbations of form (\ref{eq4}) (without losing any generality, we put the background flow value to be zero, i.e. $q_i^\ast = 0$  \citep{Kozlov_1995, Izrailsky_et_al_2004}). Hence, by setting boundary conditions ${\left. {{\Phi _i}} \right|_{{\bf{r}} \to \infty }} = 0$, and ${\left. {\partial {\Phi _i}/\partial {\bf{r}}} \right|_{{\bf{r}} \to \infty }} = 0$ to Laplace and Helmholtz equations (\ref{eq9}), we obtain two sets of Green's function superpositions satisfying system (\ref{eq9}) for the upper ($m=1$) and middle ($m=2$) layer monopole propagation cases,
\begin{eqnarray}
\label{eq10}
{\Phi _{1m}} &=& \frac{f}{\gamma }\left( \frac{{{{\left( { - 1} \right)}^n}\left( {{\alpha _n} - {\beta _n}} \right){\mu _1}}}{{{H_1}}}\log \left( {r_{i1}^\ast} \right) +\right.\\ \nonumber
 && +\left. \frac{{\left( {{\alpha _1}{\beta _2} - {\alpha _2}{\beta _1}} \right)\tau }}{{{H_3}}}\log \left( {{r_i}} \right) \right),\\ \nonumber
{\Phi _{2m}} &=&  - \frac{f}{\gamma }\left( \frac{{{{\left( { - 1} \right)}^n}\left( {{\beta _n} - 1} \right){\mu _1}}}{{{H_1}}}{K_0}\left( {\sqrt {{k_3}\left( {{\alpha _2} - 1} \right)} r_{i1}^\ast} \right) +\right.\\ \nonumber
 && +\left. \frac{{\left( {{\beta _1} - {\beta _2}} \right)\tau }}{{{H_3}}}{K_0}\left( {\sqrt {{k_3}\left( {{\alpha _2} - 1} \right)} {r_i}} \right) \right),\\ \nonumber
{\Phi _{3m}} &=&  - \frac{f}{\gamma }\left( \frac{{{{\left( { - 1} \right)}^n}\left( {1 - {\alpha _n}} \right){\mu _1}}}{{{H_1}}}{K_0}\left( {\sqrt {{k_3}\left( {{\beta _2} - 1} \right)} r_{i1}^\ast} \right) + \right.\\ \nonumber
 && +\left. \frac{{\left( {{\alpha _2} - {\alpha _1}} \right)\tau }}{{{H_3}}}{K_0}\left( {\sqrt {{k_3}\left( {{\beta _2} - 1} \right)} {r_i}} \right) \right),
\end{eqnarray}
where ${r_i} = \sqrt {{x_i}^2 + {y_i}^2} $, $r_{im}^* = \sqrt {{{\left( {{x_i} - x_m^*} \right)}^2} + {{\left( {{y_i} - y_m^*} \right)}^2}}$, and $m,\; n = 1,2$, $m \ne n$. 

Now, introducing a nonvortical plane boundary source flux in the form, $-Uy$, which does not generate any vorticity and is compensated by an analogous drain flux \citep[see a detailed substantiation in, e.g.,][]{Izrailsky_et_al_2004}, where $U$ is a characteristic velocity, one can formulate the final stream-functions of the three-layer model with the monopole moving within the $m$-layer,
\begin{equation}
\label{eq11}
{\psi _{im}} =  - Uy + {\Phi _{1m}} + {\alpha _i}{\Phi _{2m}} + {\beta _i}{\Phi _{3m}},
\end{equation}
where ${\alpha _3} = {\beta _3} = 1$. So, further we will make use of stream-functions (\ref{eq11}) with the (\ref{eq10}) set of functions $\Phi _{im}$ either for the upper ($m=1$) or middle ($m=2$) layer monopole propagation case.

Now, we can introduce certain dimensionless values, which will be used further as parameters governing the different regimes of Lagrangian advection. Introduce length scale $L = {\left( {{k_3}\left( {{\alpha _2} - 1} \right)} \right)^{ - 1/2}}$; velocity scale $U$; the Rossby number, $\varepsilon  = \frac{U}{{fL}}$; and an effective volume of the topography as $\tau  = \pi {h_0}{L^2}$, where $h_0, \;\; L$ are the height and radius of an corresponding cylinder \citep{Sokolovskiy_et_al_1998}. Then we introduce the following governing parameters,
\begin{equation}
\label{eq12}
\chi  = \frac{{f\tau }}{{{H_3}UL}} = \frac{{{h_0}\pi }}{{\varepsilon {H_3}}},\;\;{\kappa _m} = \frac{{f{\mu _m}}}{{{H_m}UL}},
\end{equation}
which characterize the dimensionless topographic vortex strength the dimensionless  monopole vortex strength, respectively. Then, by satisfying the quasi-geostrophic requirement of $\frac{h_0}{H_3} \sim O\left( \varepsilon  \right)$, we set $\chi  = \pi $. Thus, choosing the following parameters, $H_1=200 \; m$, $H_2=400 \; m$, $H_3=3000 \, m$, $\rho_1=1026.56 \; kg/m^3$, $\rho_2=1027.84 \; kg/m^3$, $\rho_3=1028.32 \; kg/m^3$, we obtain the characteristic horizontal topographic vortex scale, $L \sim 1.3 \cdot 10^4 \; m$.
\section{Equations of motion}
Now, by making use of the dimensionless parameters and the geostrophic relations, one can write the equations of motion for the monopole's center and for a fluid particle, being advected by the monopole-topography interaction velocity field. The monopole motion in the $m$-layer is governed by the following equations of motion,
\begin{eqnarray}
\label{eq13}
\frac{d}{{dt}}x_m^* &=& {\left. { - \frac{{\partial {\psi _{mm}}}}{{\partial y}}} \right|_{\scriptstyle x = x_m^*\hfill\atop
\scriptstyle y = y_m^*\hfill}} = W + \chi \frac{{y_m^*}}{{r_m^*}}{V_m}\left( {r_m^*} \right),\\ \nonumber
\frac{d}{{dt}}y_m^* &=& {\left. {\frac{{\partial {\psi _{mm}}}}{{\partial x}}} \right|_{\scriptstyle x = x_m^*\hfill\atop
\scriptstyle y = y_m^*\hfill}} = - \chi \frac{{x_m^*}}{{r_m^*}}{V_m}\left( {r_m^*} \right),
\end{eqnarray}
where $m=1,2$, and $W=W(t)$ is the dimensionless background flow velocity; 
\begin{eqnarray*}
&&{V_m}\left( \xi  \right) = \frac{1}{\gamma }\left( \left( {{\alpha _1}{\beta _2} - {\alpha _2}{\beta _1}} \right)\frac{1}{\xi } + {\alpha _m}\left( {{\beta _1} - {\beta _2}} \right){K_1}\left( \xi  \right) +\right.\\
&&\left. + {\beta _m}\left( {{\alpha _2} - {\alpha _1}} \right)\sqrt {\frac{{\left( {{\beta _2} - 1} \right)}}{{\left( {{\alpha _2} - 1} \right)}}} {K_1}\left( {\sqrt {\frac{{\left( {{\beta _2} - 1} \right)}}{{\left( {{\alpha _2} - 1} \right)}}} \xi } \right) \right),\\ \nonumber
\end{eqnarray*} 
and $r_m^* = \sqrt {{{\left( {x_m^*} \right)}^2} + {{\left( {y_m^*} \right)}^2}} $ is the monopole position with $m=1,2$ for the upper- and middle-layer monopole motion, respectively. System (\ref{eq13}) is an elaborated system to govern the dynamics of a fluid particle due to the velocity field being generated by an exterior background flow intersecting a delta function bottom irregularity. Lagrangian advection being determined by system (\ref{eq13}) has been studied recently in the frame of barotropic \citep{Sokolovskiy_et_al_1998, Izrailsky_et_al_2004, Koshel_Prants_2006}, two-layer \citep{Kozlov_et_al_2005, Ryzhov_Koshel_2011b} and three-layer baroclinic geophysical flows \citep{Ryzhov_Koshel_2011a}. In our case, however, system (\ref{eq13}) governs not a fluid particle's motion, but a singular vortex's center motion. So, the upper- and middle-layer monopoles themselves move as fluid particles due to the topographic vortex velocity field. Fluid particles of the monopole-topography interaction system, although, undergo the joint influence of both the monopole and topographic vortex velocity fields.

Motion of a fluid particle being influenced by the cooperate monopole-topography velocity field obeys to the relations,
\begin{eqnarray}
\label{eq14}
{{\dot x}_i} &=&  - \frac{{\partial {\psi _{im}}}}{{\partial {y_i}}} =\\ \nonumber
&=& W + {\kappa _m}\frac{{\left( {{y_i} - y_m^*} \right)}}{{r_{im}^*}}{P_{im}}\left( {r_{im}^*} \right) + \chi \frac{{{y_i}}}{{{r_i}}}{V_i}\left( {{r_i}} \right),\\ \nonumber
{{\dot y}_i} &=& \frac{{\partial {\psi _{im}}}}{{\partial {x_i}}} = \\ \nonumber
&=& - \left( {{\kappa _m}\frac{{\left( {{x_i} - x_m^*} \right)}}{{r_{im}^*}}{P_{im}}\left( {r_{im}^*} \right) + \chi \frac{{{x_i}}}{{{r_i}}}{V_i}\left( {{r_i}} \right)} \right),
\end{eqnarray}
where $r_{im}^* = \sqrt {{{\left( {{x_i} - x_m^*} \right)}^2} + {{\left( {{y_i} - y_m^*} \right)}^2}} $ is the fluid particle position relatively to the monopole's center position, 
\begin{eqnarray*}
&&P_{im}\left( \xi  \right) = \frac{\left( - 1\right)^n}{\gamma }\left( \left( \alpha _n - \beta _n \right)\frac{1}{\xi } \right. \left. + \alpha _i\left( \beta _n - 1 \right){K_1}\left( \xi  \right) +\right.\\
&&\left. + \beta _i\left( 1 - \alpha _n \right)\sqrt {\frac{\left( \beta _2 - 1 \right)}{\left( \alpha _2 - 1 \right)}} {K_1}\left( \sqrt {\frac{\left( \beta _2 - 1 \right)}{\left( \alpha _2 - 1 \right)}} \xi \right) \right), \nonumber
\end{eqnarray*} and $m, \; n=1,2$, $m \ne n$.
\section{Monopole motion}
\begin{figure}[t]
\includegraphics[width=8.3cm]{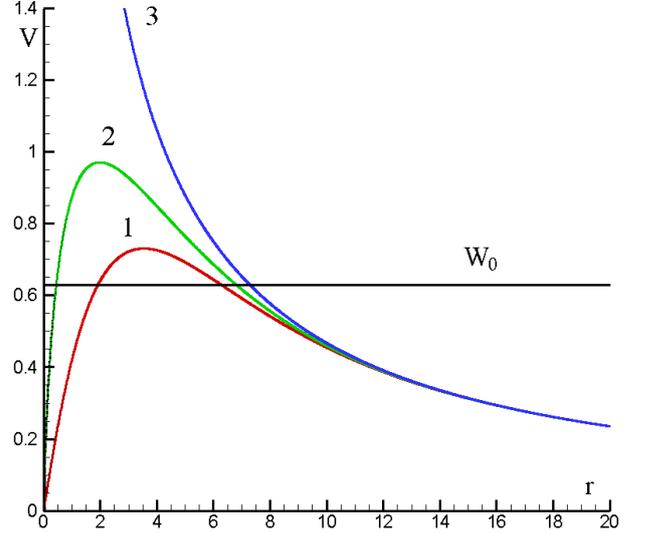}
\caption{Azimuthal velocities of the topographic vortex within the layers. Curves 1,2,3 correspond to the upper, middle, and bottom layer, respectively. The horizontal straight line indicates constant background velocity value $W_0=0.2\pi$.}
\label{fig1a}
\end{figure}

\begin{figure}[t]
\includegraphics[width=8.3cm]{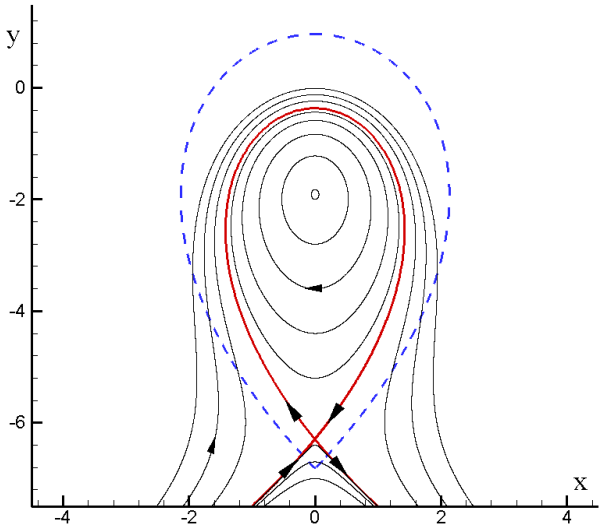}
\caption{Topographic vortex stream-lines for the upper layer. The red curve indicates the separatrix. The dashed blue curve corresponds to the middle layer separatrix.}
\label{fig1b}
\end{figure}

First, we briefly analyze system (\ref{eq13}). that governs the monopole's dynamics. An elaborated study of this system has been conducted in \citep{Ryzhov_Koshel_2011a}. If the background exterior flow is constant ($W=W_0$), system (\ref{eq13}) is integrable in the sense of the stream-line-trajectory coincidence \citep[e.g.,][]{Zaslavsky_1998}. Due to the bottom topography is singular, any nonzero value of $W_0$ always produces a closed Taylor column region called a topographic vortex within the lower layer. To have such closed regions within the middle and upper layers, however, the background velocity should be lower than a critical value. This critical value is the maximal value of the azimuthal velocity in the corresponding layer. So, if one chooses the background flow to satisfy this condition, then three different-size Taylor columns will occur due to the bottom irregularity. These three columns may be thought of as a discrete Taylor cone. Figure \ref{fig1a} depicts azimuthal velocities $V_i$ depending on distance $r$ to the topographic vortex elliptic point. We chose $W_0=0.2,\;\chi=0.2\pi$ to ensure the mesoscale closed regions to exist in the all three layers. The points, where the horizontal line intersects the azimuthal velocity curves, correspond to elliptic and hyperbolic critical points of the vortex. Figure \ref{fig1b} demonstrates stream-lines of the resulting topographic vortex in the upper-layer. The red curve indicates the separatrix dividing the flow into the vortical region and the exterior flow. Since we also are interested in the middle-layer monopole propagation case, the vortical region of the middle layer is indicated by the blue dashed curve. 

\section{Fluid particle advection}
\subsection{Regular monopole motion}
\begin{figure*}[t]
\vspace*{2mm}
\begin{center}
\begin{subfigure}[b]{0.2\textwidth}
\includegraphics[width=4.0cm]{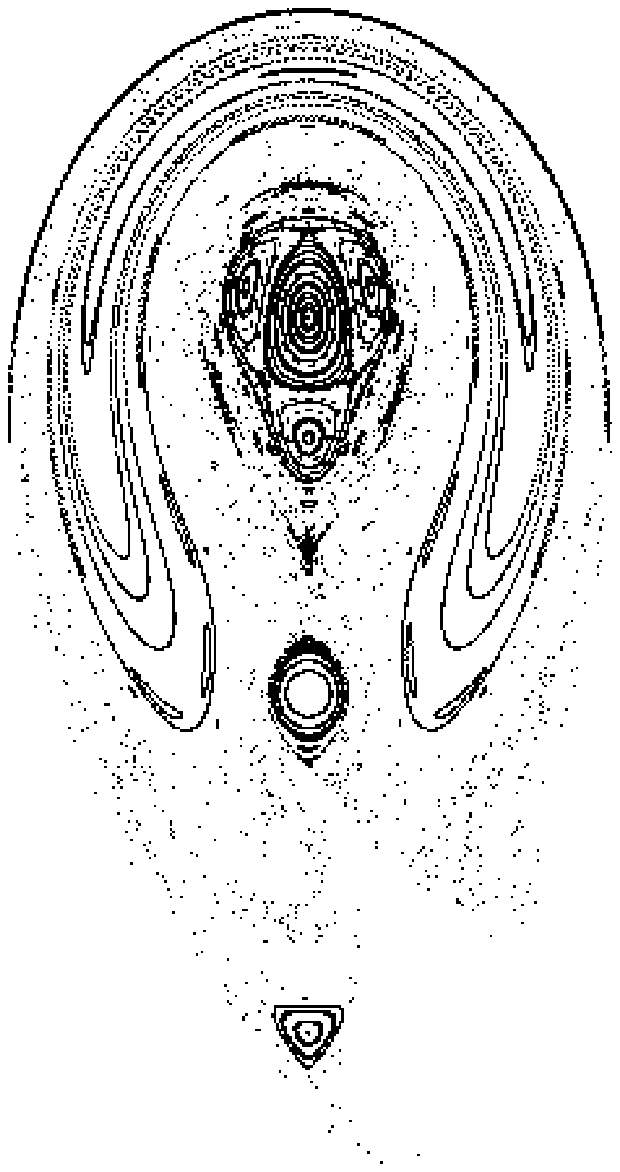}
\caption{$\left(0,\; -4\right)$} 
\end{subfigure}
\begin{subfigure}[b]{0.2\textwidth}
\includegraphics[width=4.0cm]{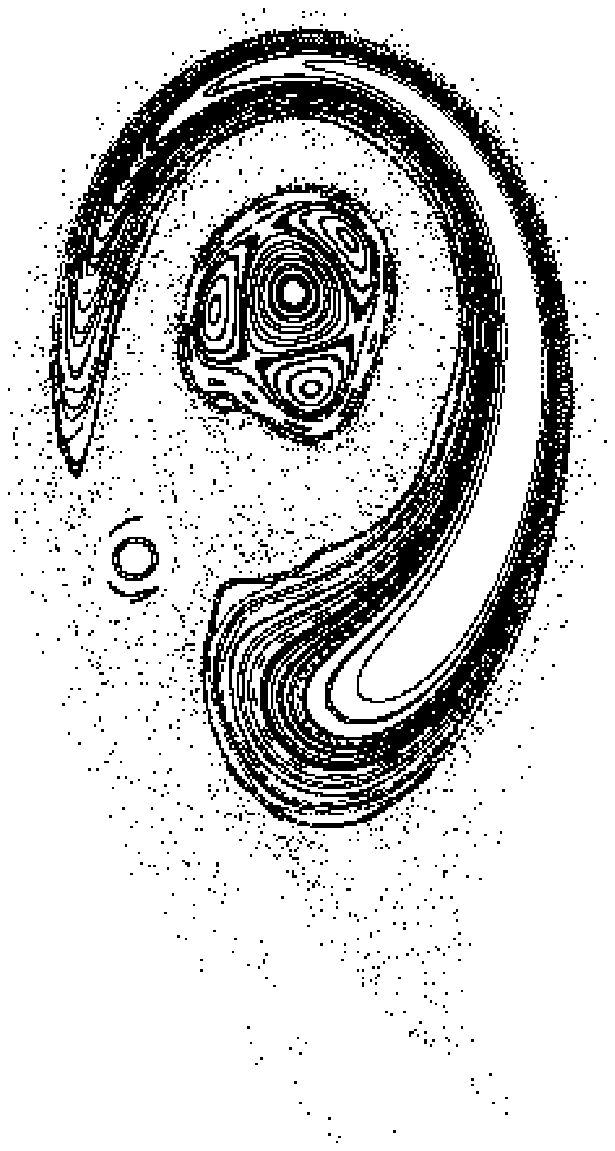}
\caption{$\left(-0.8,\; -3.3107\right)$} 
\end{subfigure}
\begin{subfigure}[b]{0.2\textwidth}
\includegraphics[width=4.0cm]{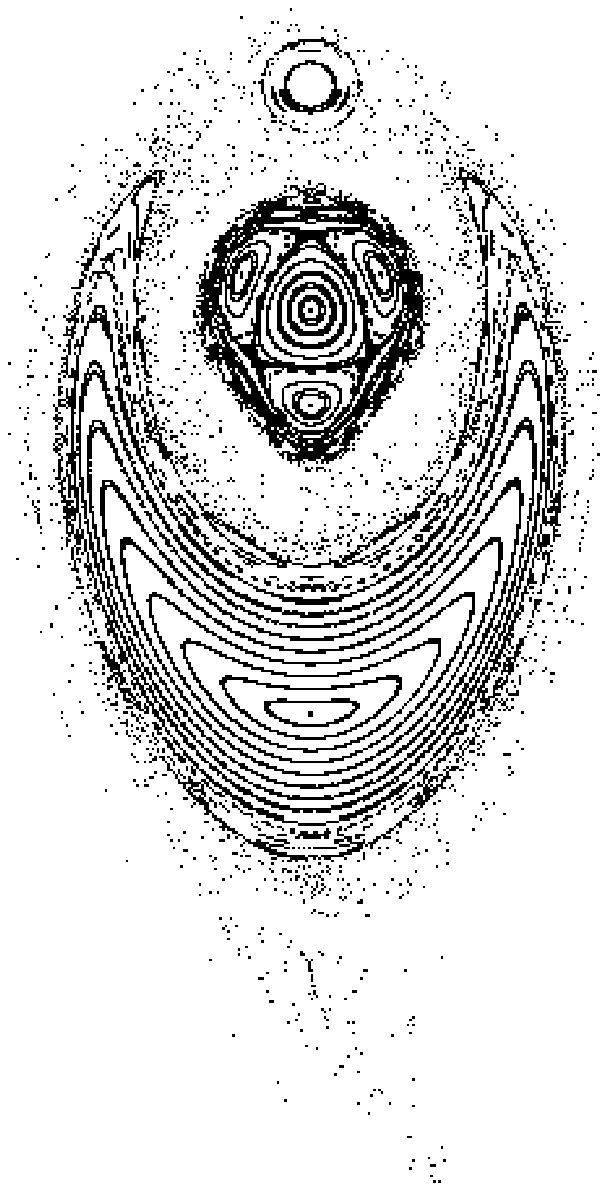}
\caption{ $\left(0,\; -0.6936\right)$}
\end{subfigure}
\end{center}
\caption{Equivalent Poincar\'{e} sections of system (\ref{eq14}) for the same values of $\kappa = 0.01$ and the same values of perturbation frequency $\omega$ but different monopole initial positions $\left(x_1^\ast,\; y_1^\ast\right)$.}
\label{fig2}
\end{figure*}

Now we can analyze fluid particle advection being induced by the monopole-topography interaction velocity field. Motion of a fluid particle is governed by system (\ref{eq14}), where the right part of the relations comprises the monopole motion solution given by (\ref{eq13}). First, we consider the periodic solution of (\ref{eq13}). This solution, although cannot be expressed in an analytical form, is time-dependent with a period being equal to the time of the monopole passing a closed trajectory within the separatrices shown in \ref{fig1b}. Hence, system (\ref{eq14}) is a dynamical system with one and a half degrees of freedom, that permits to occur the fluid particle irregular dynamics which is conventionally called chaotic advection \citep{Aref_1984, Wiggins_1992, Aref_2002}. Chaotic advection manifests itself through exponential divergence of close trajectories in a finite time \citep[e.g.,][]{Lichtenberg_Lieberman_1983, Zaslavsky_1998}. The easiest way to demonstrate the chaotic advection manifestation is by constructing Poincar\'{e} sections of system (\ref{eq14}). Figure \ref{fig2}a shows a Poincar\'{e} section as $\kappa _1 = 0.01$, $y_1^\ast \left(0\right) = -4$, corresponding to frequency $\omega = 0.1611$ of monopole rotation along an orbit shown in fig. \ref{fig1b}.

That half degree of freedom corresponds to a time-dependent perturbation, which concerning system (\ref{eq14}) is the monopole motion term comprising strength $\kappa _m$. However, this monopole strength is not the only parameter greatly affecting Lagrangian advection, the initial position of the monopole is also of great importance. As the initial position parameter, we choose the positions on the $y$-axis due to all the stream-lines shown in fig. \ref{fig1b} intersect this line. Figures \ref{fig2}b,c show the system (\ref{eq14}) phase space equivalent structures for different monopole initial positions, nevertheless, corresponding to the same stream-line. Hence, the positions on the $y$-axis correspond to all the frequencies of the monopole rotation about topography. Thus, further we will address how the monopole's strength and initial position parameters affect the fluid particle dynamics. 

The Poincar\'{e} section analysis is a very useful technique to estimate which part of fluid particles is involved either in regular advection or in chaotic advection, however, this technique fails to show what happens with fluid particles in certain moment of time. So, to address the question, how these fluid particles move during the monopole passing a revolution about the topography, we calculate the number of the flow's critical points that appear at each instant  \citep{Ryzhov_Koshel_2011b, Ryzhov_et_al_2012}. The simple idea of this classification is that the more critical points of their initial set survive or, in other worlds, the less topological changes appear during a monopole revolution the more regular system (\ref{eq14}) is.
\subsection{Diagram of the number of the critical points}
\begin{figure}[t]
\vspace*{2mm}
\begin{center}
\begin{subfigure}[t]{0.2\textwidth}
\includegraphics[width=3.5cm]{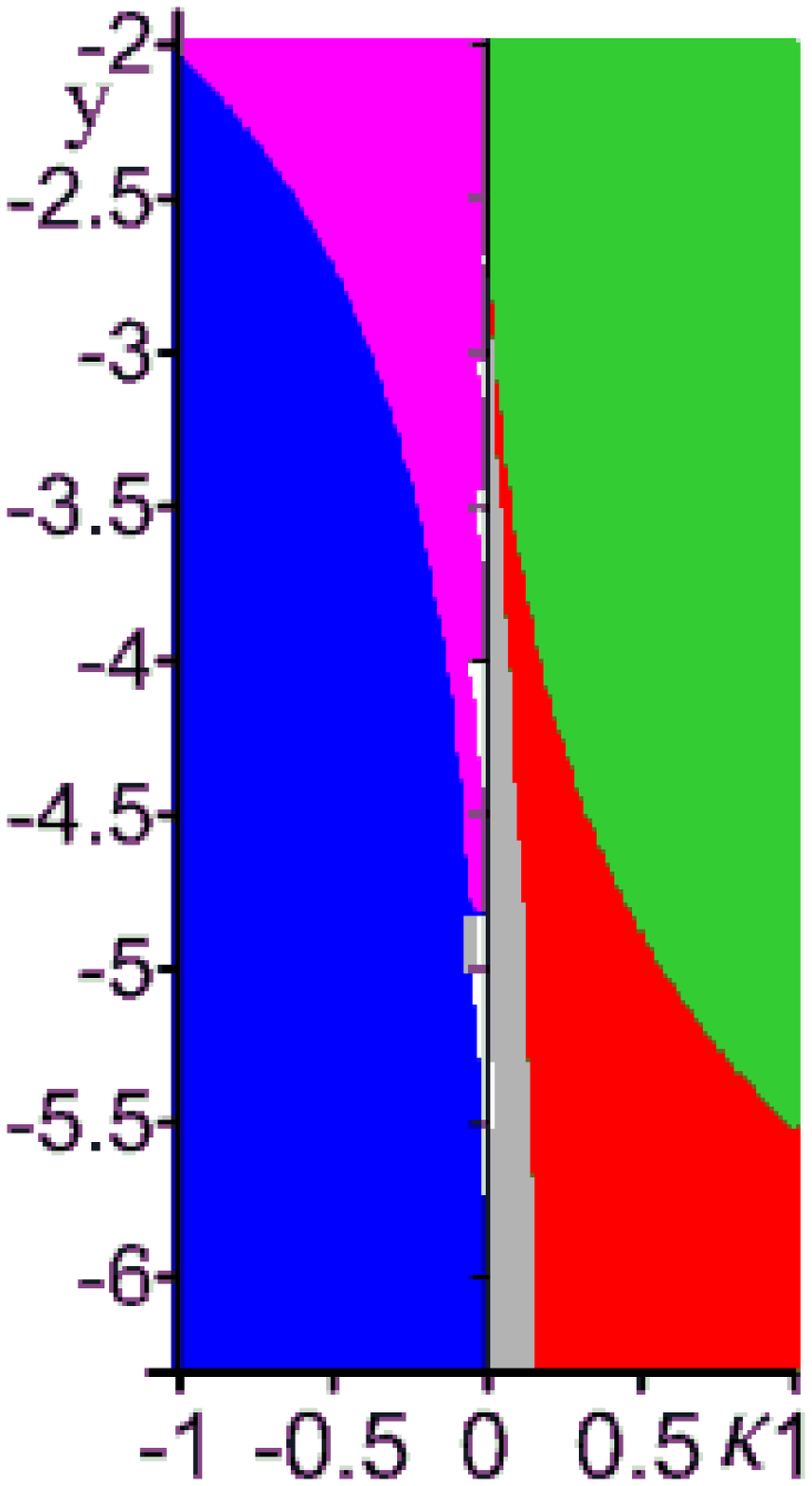}
\caption{upper-layer monopole propagation case} 
\end{subfigure}\quad
\begin{subfigure}[t]{0.2\textwidth}
\includegraphics[width=3.5cm]{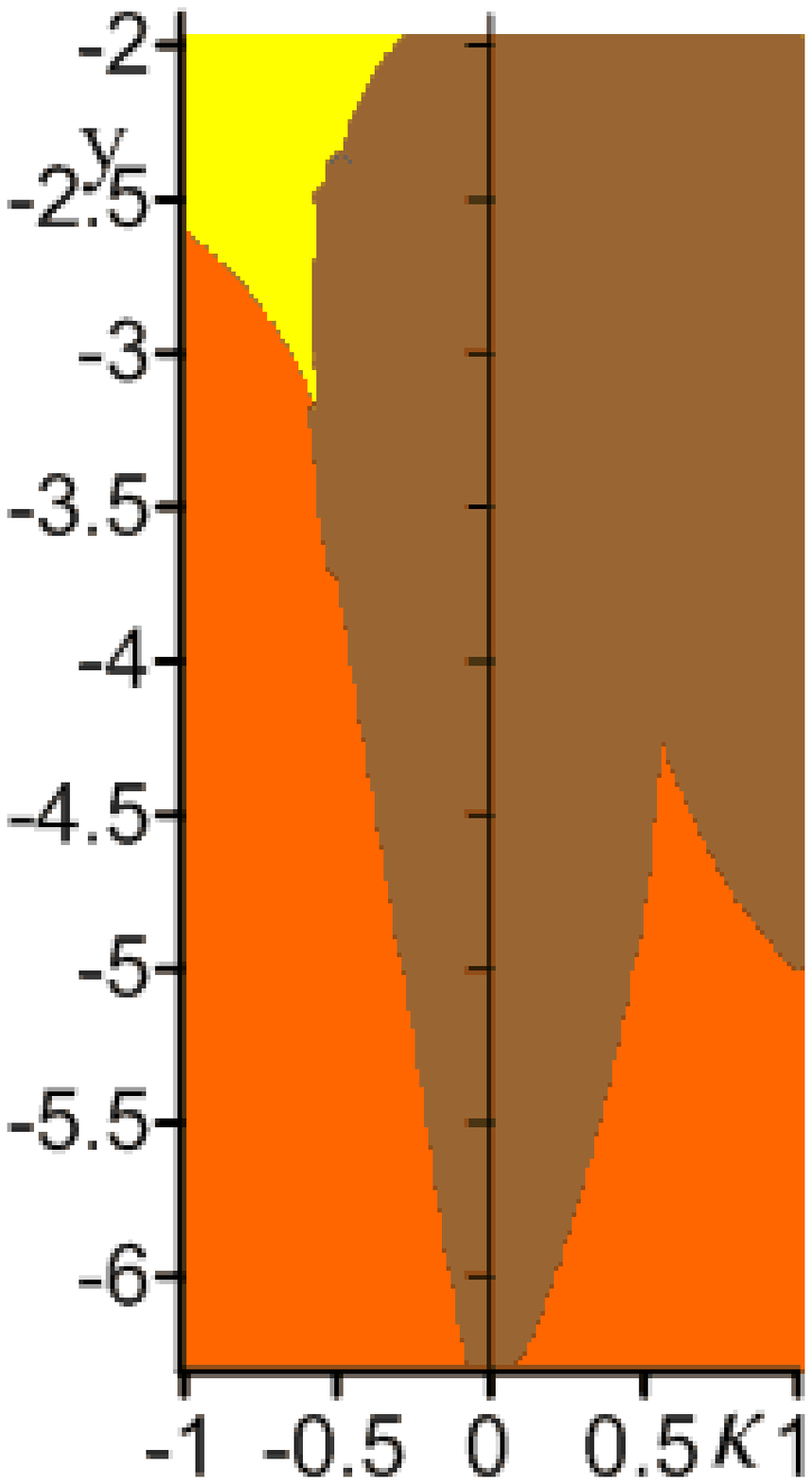}
\caption{middle-layer monopole propagation case} 
\end{subfigure}
\end{center}
\caption{Number of the flow's regular critical points by color. Blue $-$ 3 initial and 3 or 5 half-period points; purple $-$ 5 initial and 5 half-period points, grey $-$ 3 initial and 3 half-period points red $-$ 3 initial and 1 half-period points; green $-$ 1 initial and 1 half-period points. Yellow $-$ 6 initial and 2 half-period points, orange $-$ 4 initial and 2 half-period points, brown $-$ 2 initial and 2 half-period points}
\label{fig3}
\end{figure}

As the monopole moves about topography, the number of the flow's regular critical points changes, that results in the flow topology altering its characteristics in time \citep[e.g.,][]{Aref_Brons_1998}. It should be mentioned, that, in the upper layer monopole propagation case, one singular critical point corresponding to the monopole's center always exists, so, we have excluded it from the consideration. Although, in the middle-layer monopole propagation case, no singular points occur within the upper-layer velocity field due to the singular middle-layer monopole appears as a regular one in the upper layer. So, making use of the introduced classification, we present diagrams of the number of the regular critical points in the upper-layer monopole propagation case and in the middle-layer monopole propagation case, respectively, depending on monopole's strength $\kappa$ and initial position $y$. These diagrams depict by color how many regular critical points appear at the beginning of the monopole rotation (initial critical points) and at the time the monopole passes a half of its rotation period (half-period critical points). Figure \ref{fig3}a, and fig. \ref{fig3}b correspond to the upper-layer monopole propagation case and to the middle-layer monopole propagation case, respectively.

\begin{figure*}[t]
\vspace*{2mm}
\begin{center}
\begin{subfigure}[t]{0.22\textwidth}
\includegraphics[width=3.5cm]{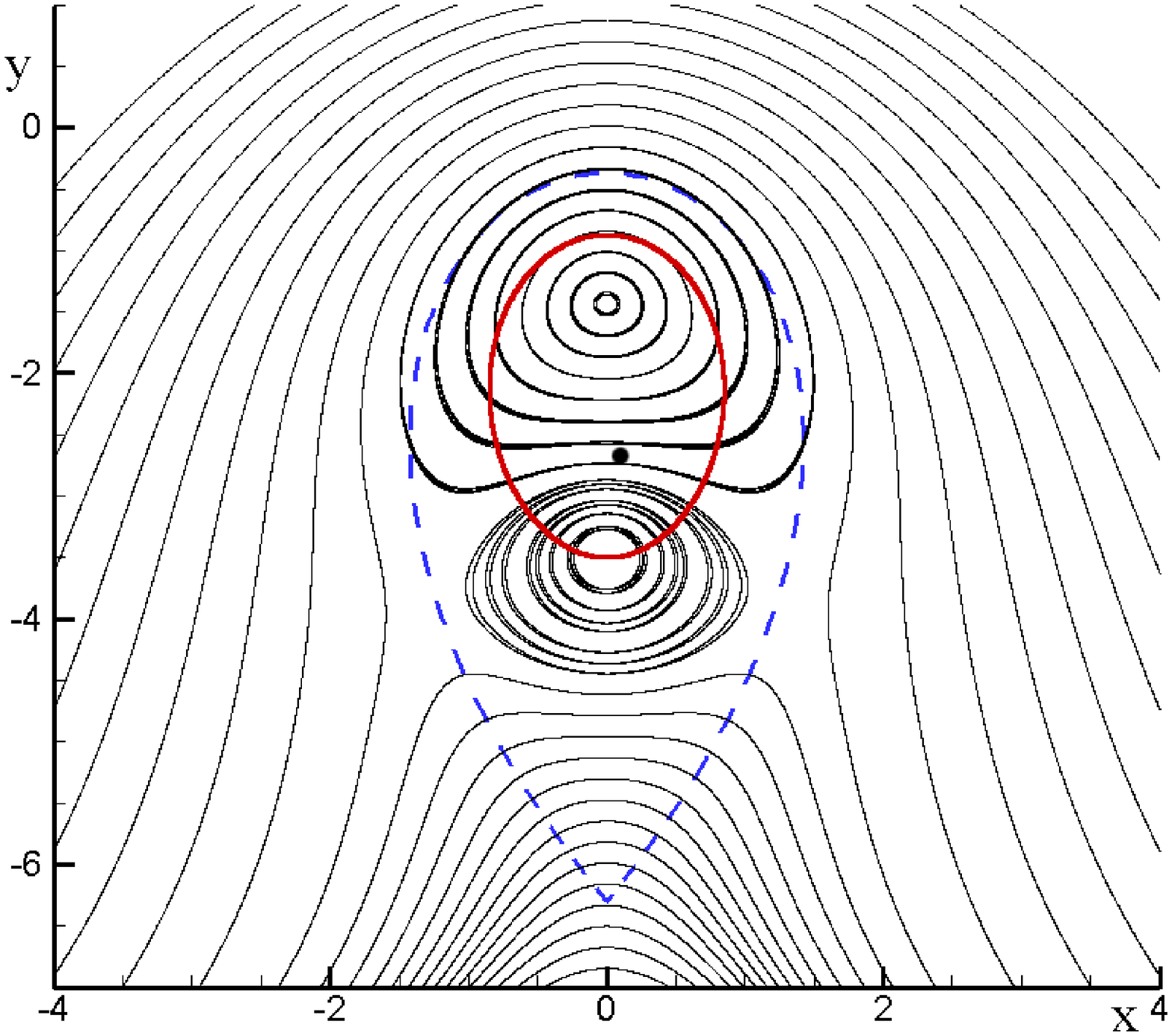}
\caption{initial stage for the blue color region $\left(\kappa=-0.5,\;y=-3.5\right)$} 
\end{subfigure}
\quad
\begin{subfigure}[t]{0.22\textwidth}
\includegraphics[width=3.5cm]{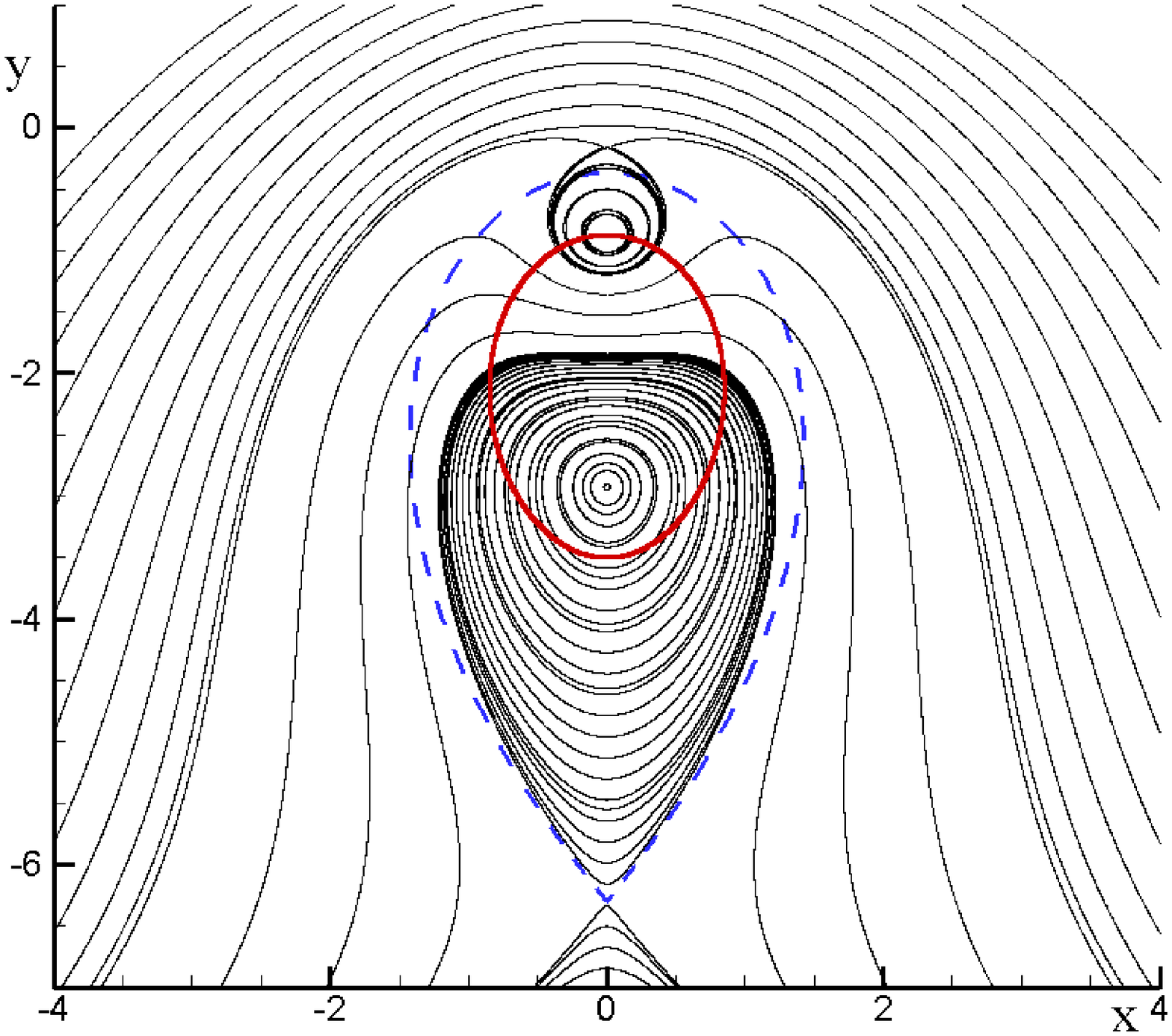}
\caption{half-period stage for the blue color region $\left(\kappa=-0.5,y=-3.5\right)$} 
\end{subfigure}
\quad
\begin{subfigure}[t]{0.22\textwidth}
\includegraphics[width=3.5cm]{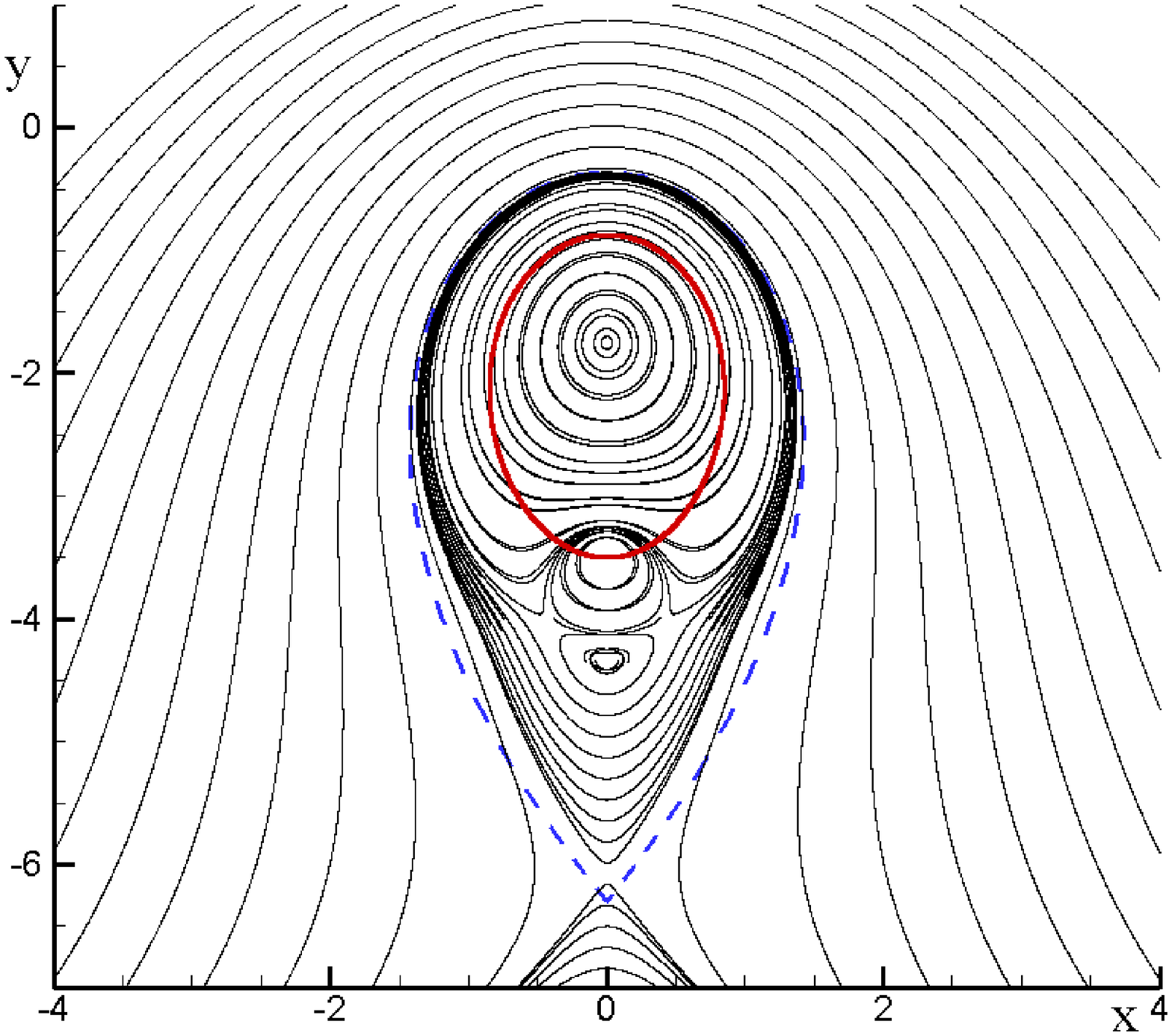}
\caption{initial stage for the purple color region $\left(\kappa=-0.1,\;y=-3.5\right)$} 
\end{subfigure}
\quad
\begin{subfigure}[t]{0.22\textwidth}
\includegraphics[width=3.5cm]{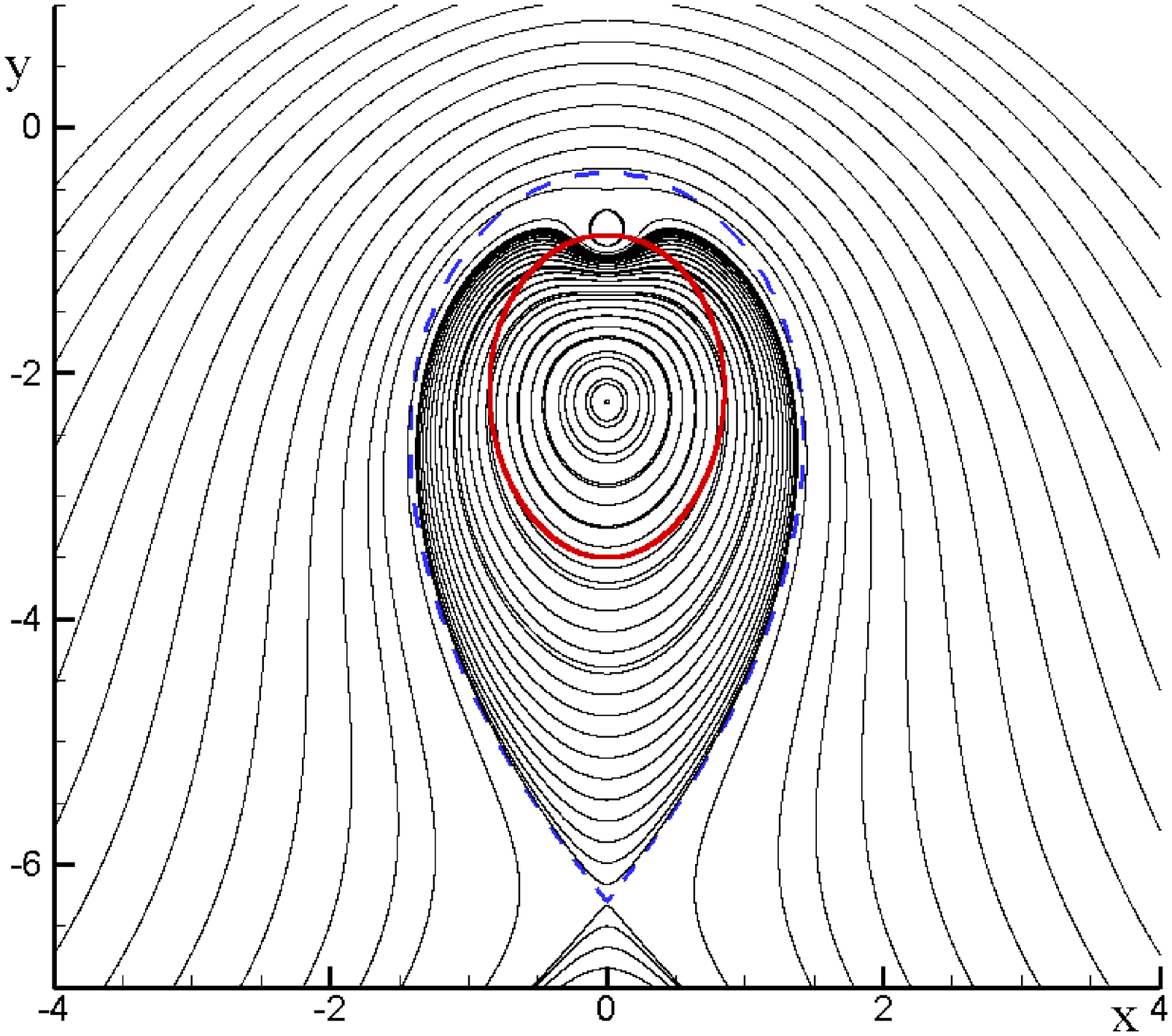}
\caption{half-period stage for the purple color region $\left(\kappa=-0.1,\;y=-3.5\right)$} 
\end{subfigure}
\begin{subfigure}[t]{0.22\textwidth}
\includegraphics[width=3.5cm]{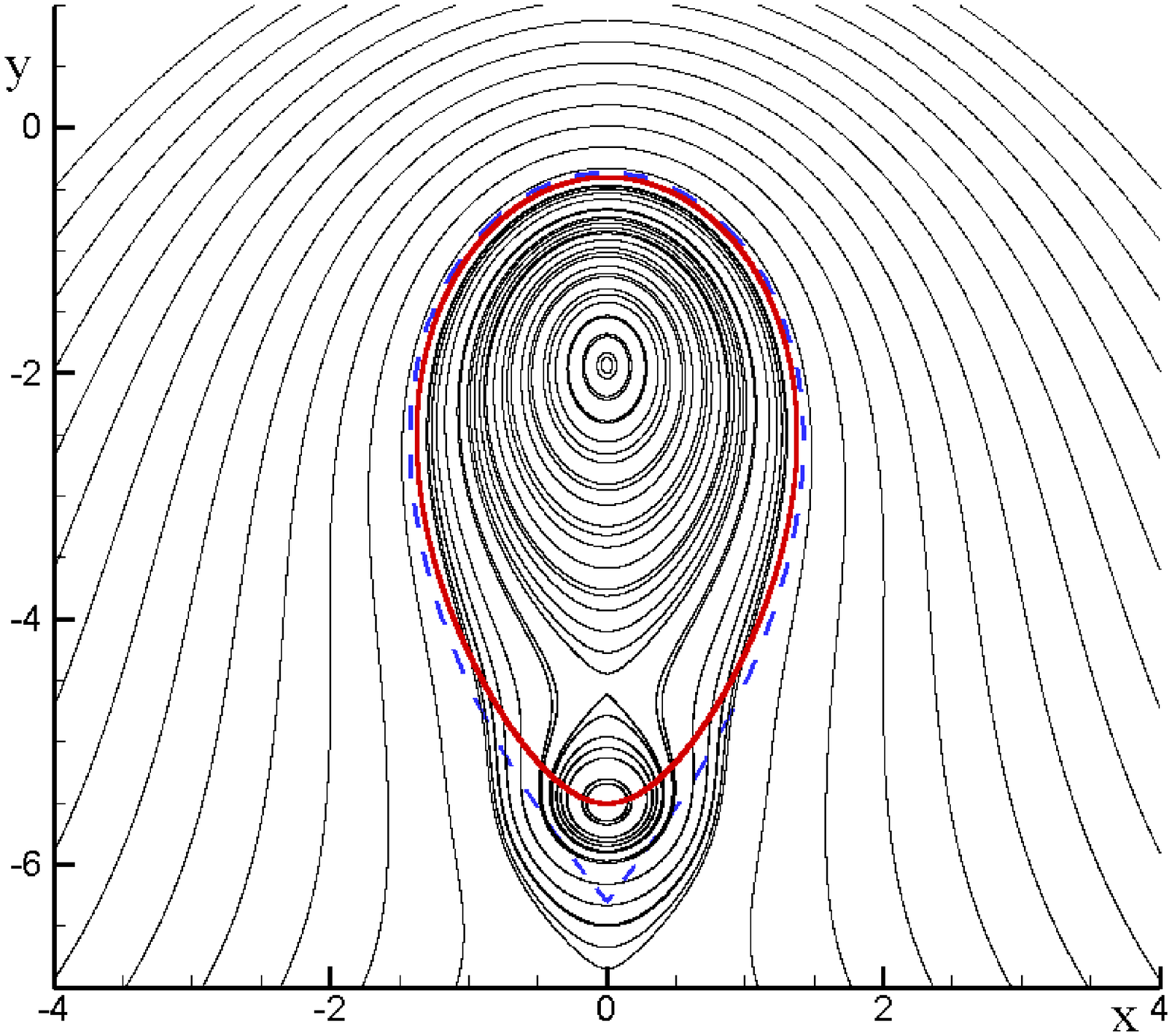}
\caption{initial stage for the grey color region $\left(\kappa=0.1,\;y=-5.5\right)$} 
\end{subfigure}
\quad
\begin{subfigure}[t]{0.22\textwidth}
\includegraphics[width=3.5cm]{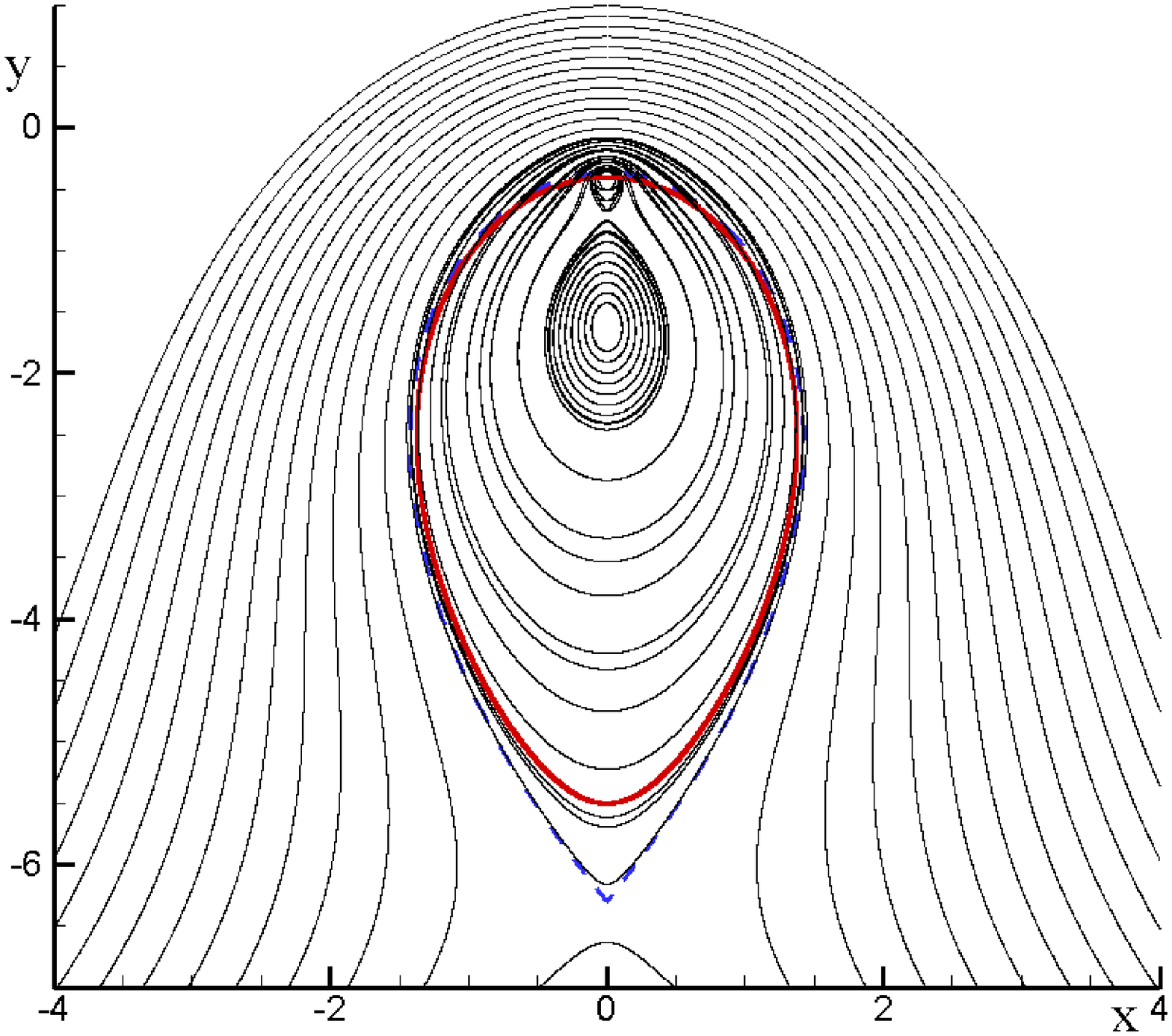}
\caption{half-period stage for the grey color region $\left(\kappa=0.1,\;y=-5.5\right)$} 
\end{subfigure}
\quad
\begin{subfigure}[t]{0.22\textwidth}
\includegraphics[width=3.5cm]{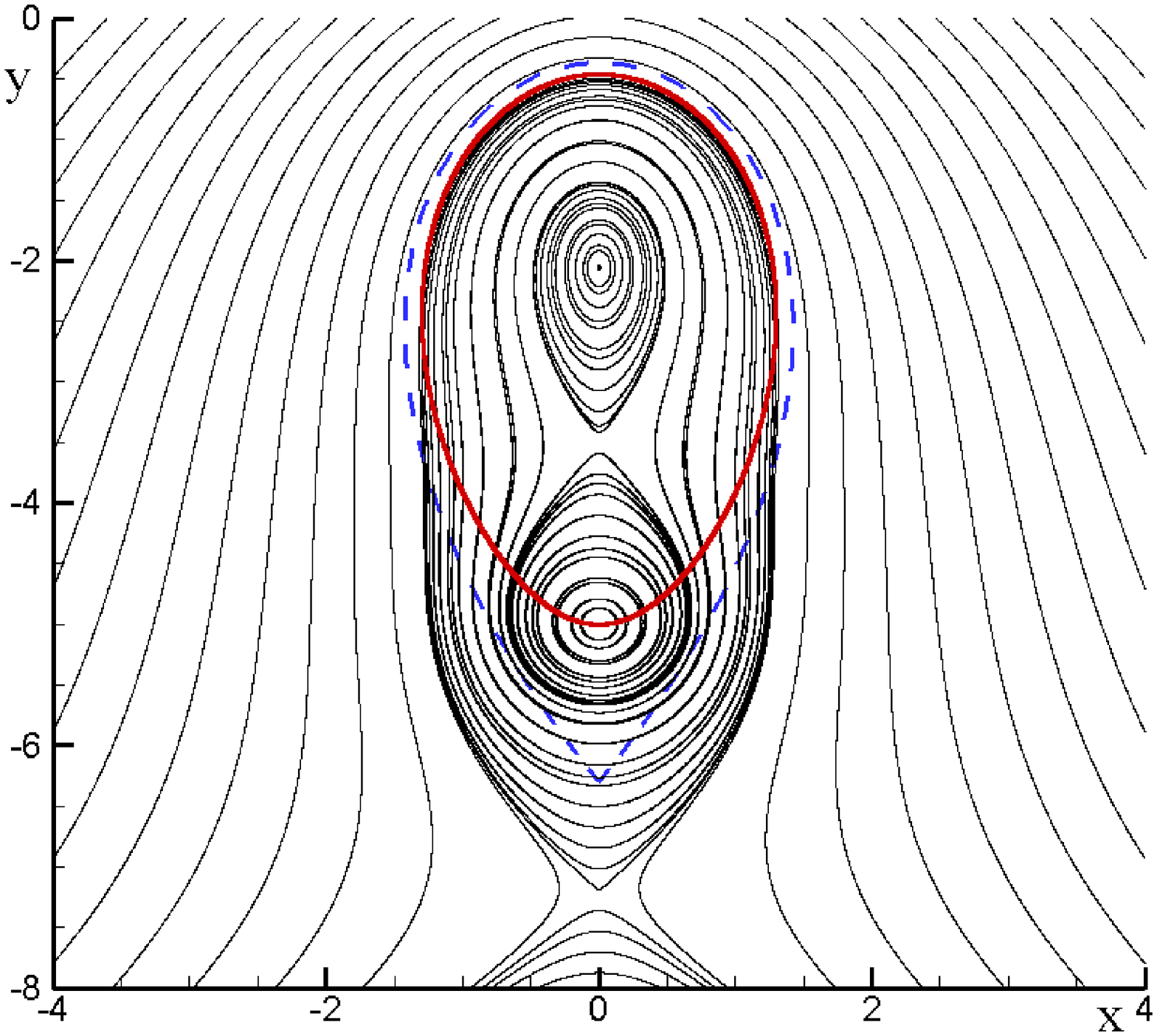}
\caption{initial stage for the red color region $\left(\kappa=0.3,\;y=-5\right)$} 
\end{subfigure}
\quad
\begin{subfigure}[t]{0.22\textwidth}
\includegraphics[width=3.5cm]{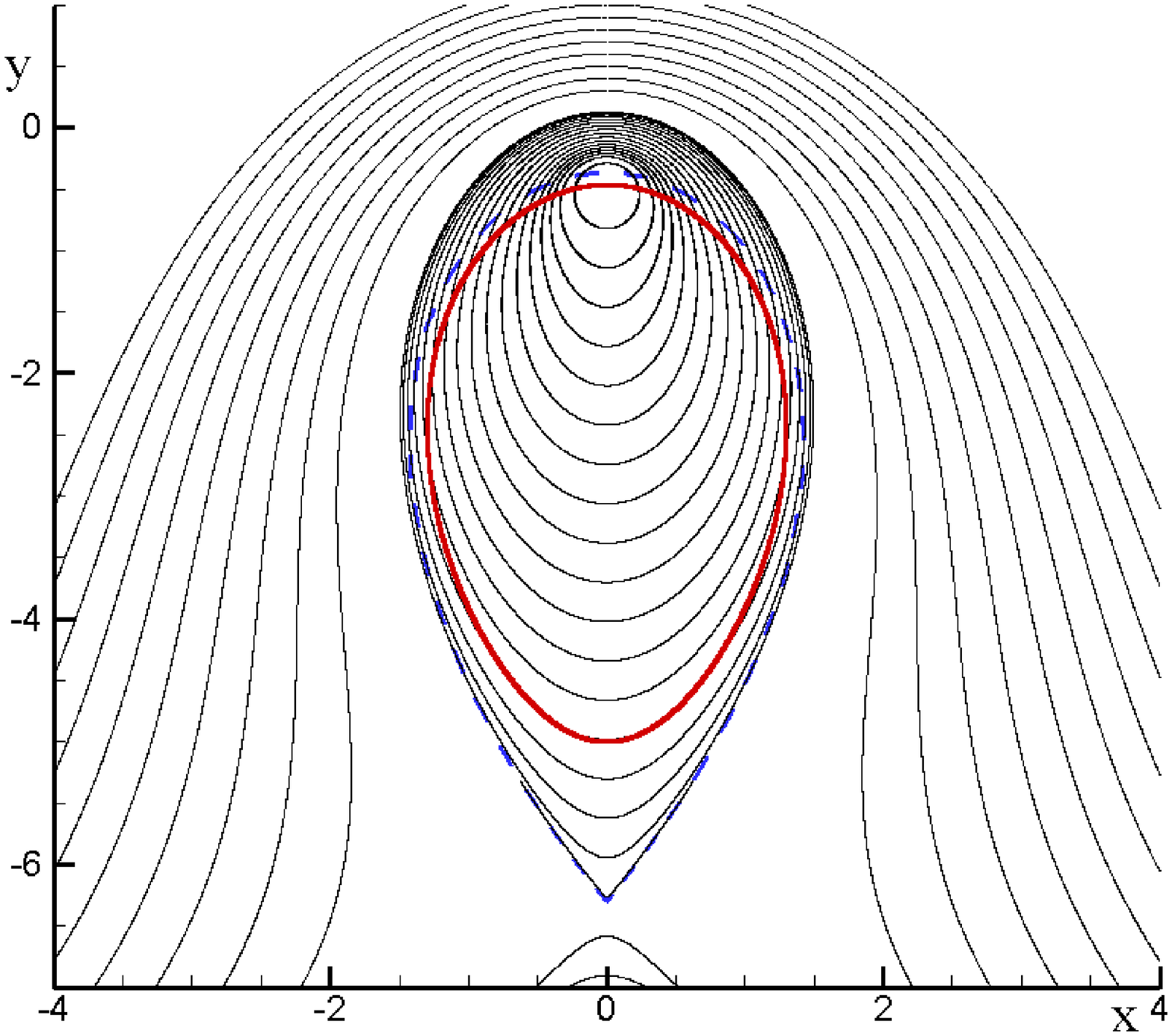}
\caption{half-period stage for the red color region $\left(\kappa=0.3,\;y=-5\right)$} 
\end{subfigure}
\quad
\begin{subfigure}[t]{0.22\textwidth}
\includegraphics[width=3.5cm]{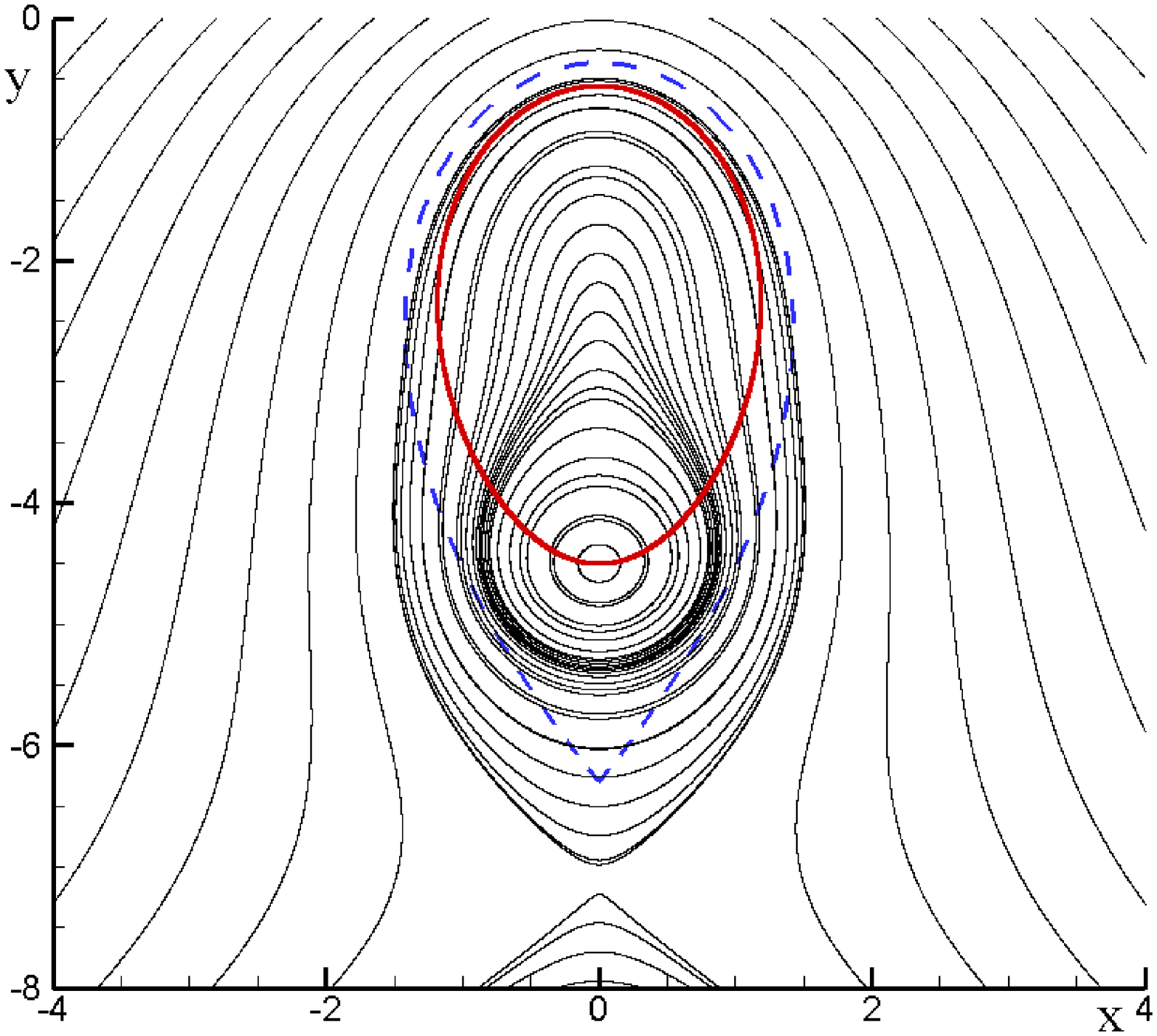}
\caption{initial stage for the green color region $\left(\kappa=0.5,y=-4.5\right)$} 
\end{subfigure}
\quad
\begin{subfigure}[t]{0.22\textwidth}
\includegraphics[width=3.5cm]{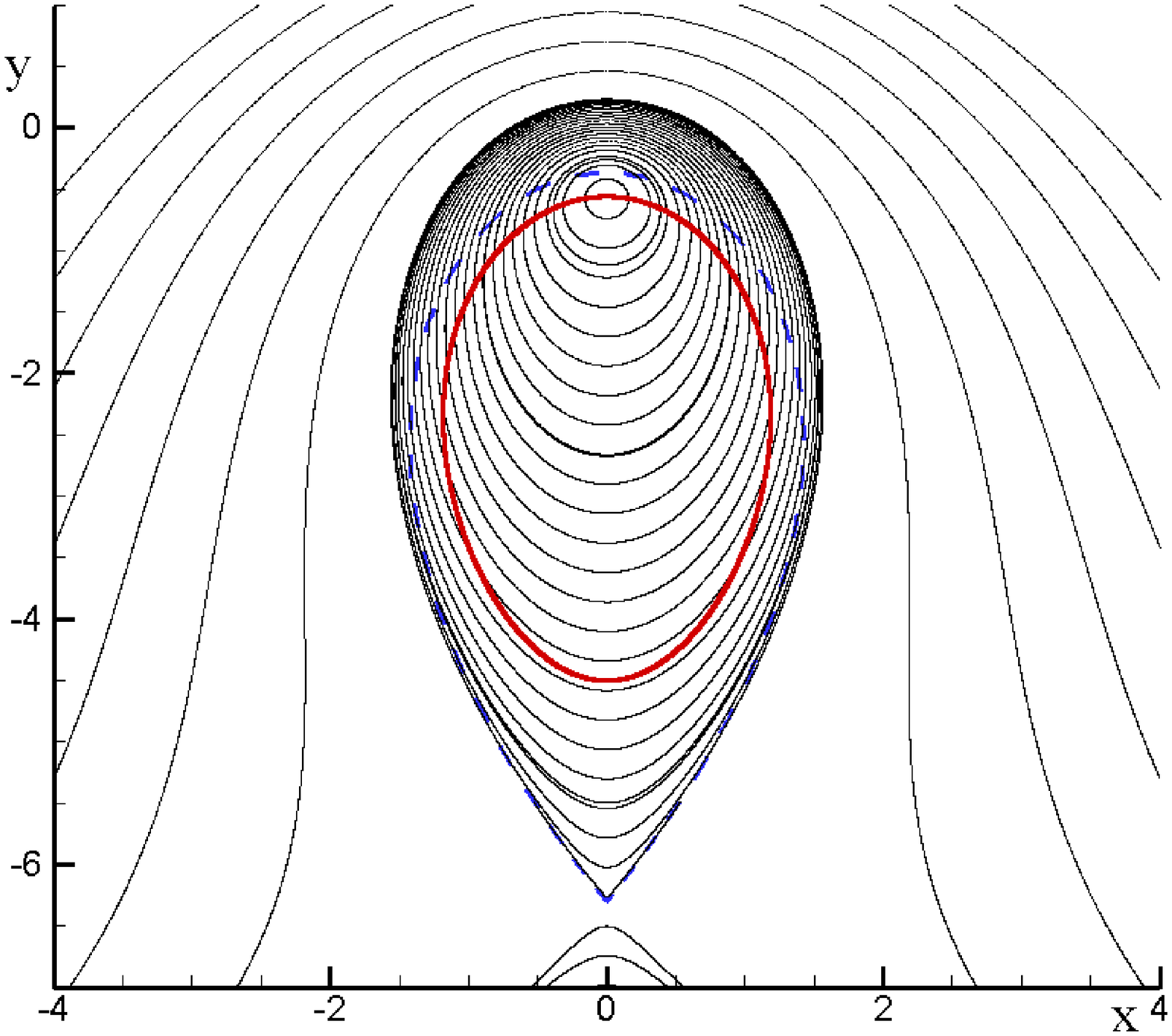}
\caption{half-period stage for the green color region $\left(\kappa=0.5,y=-4.5\right)$} 
\end{subfigure}
\end{center}
\caption{Stream-lines of the flow in the upper-layer monopole propagation case. Red curve corresponds to the monopole motion trajectory. Dashed blue curve is the topographic vortex unperturbed separatrix.}
\label{fig4}
\end{figure*}

Now, we offer a detailed explanation for the diagrams. $\kappa<$0 region corresponds to counter-rotation of the monopole and topographic vortex, and $\kappa>0$ corresponds to co-rotation of the monopole and topographic vortex. First, we consider fig. \ref{fig3}a depicting the diagram associated with the upper-layer monopole propagation case. Figure \ref{fig4} shows the flow's stream-lines at the initial stage of monopole motion and at the half-period stage. The red curves are the monopole trajectories, and the dashed blue curve corresponds to the unperturbed topographic vortex separatrix. 

\begin{figure*}[t]
\vspace*{2mm}
\begin{center}
\begin{subfigure}[t]{0.14\textwidth}
\includegraphics[width=2.5cm]{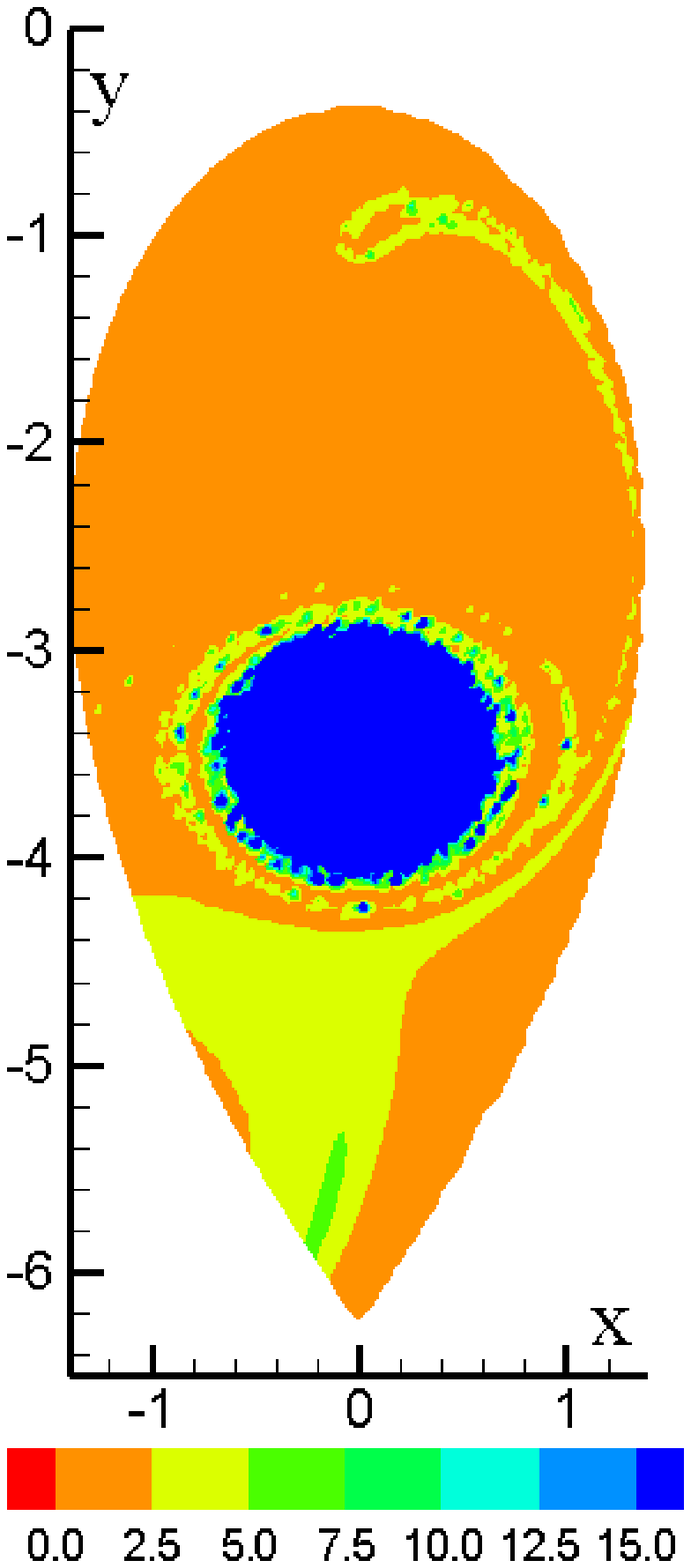}
\caption{$\kappa=-0.5$,\\ $y=-3.5$} 
\end{subfigure}
\quad
\begin{subfigure}[t]{0.14\textwidth}
\includegraphics[width=2.5cm]{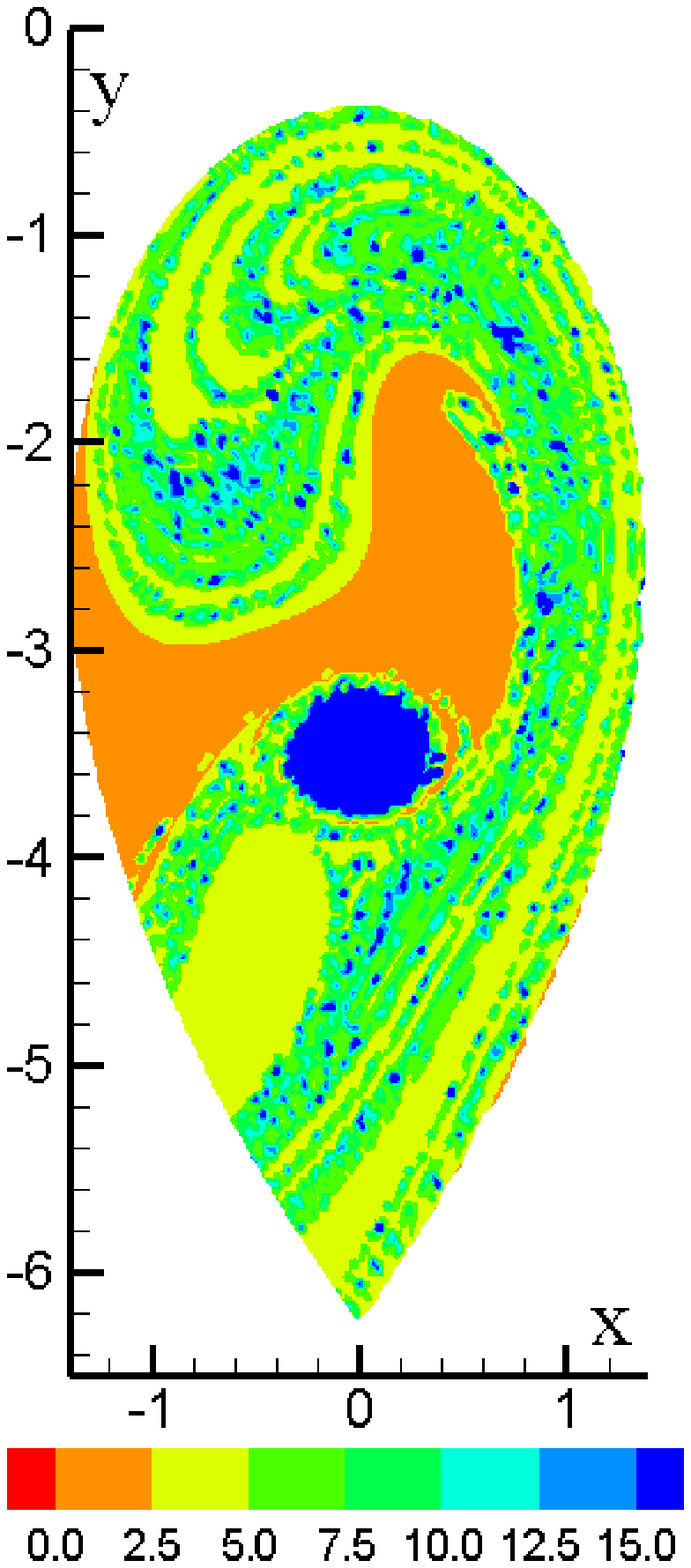}
\caption{$\kappa=-0.1$,\\ $y=-3.5$} 
\end{subfigure}
\quad
\begin{subfigure}[t]{0.14\textwidth}
\includegraphics[width=2.5cm]{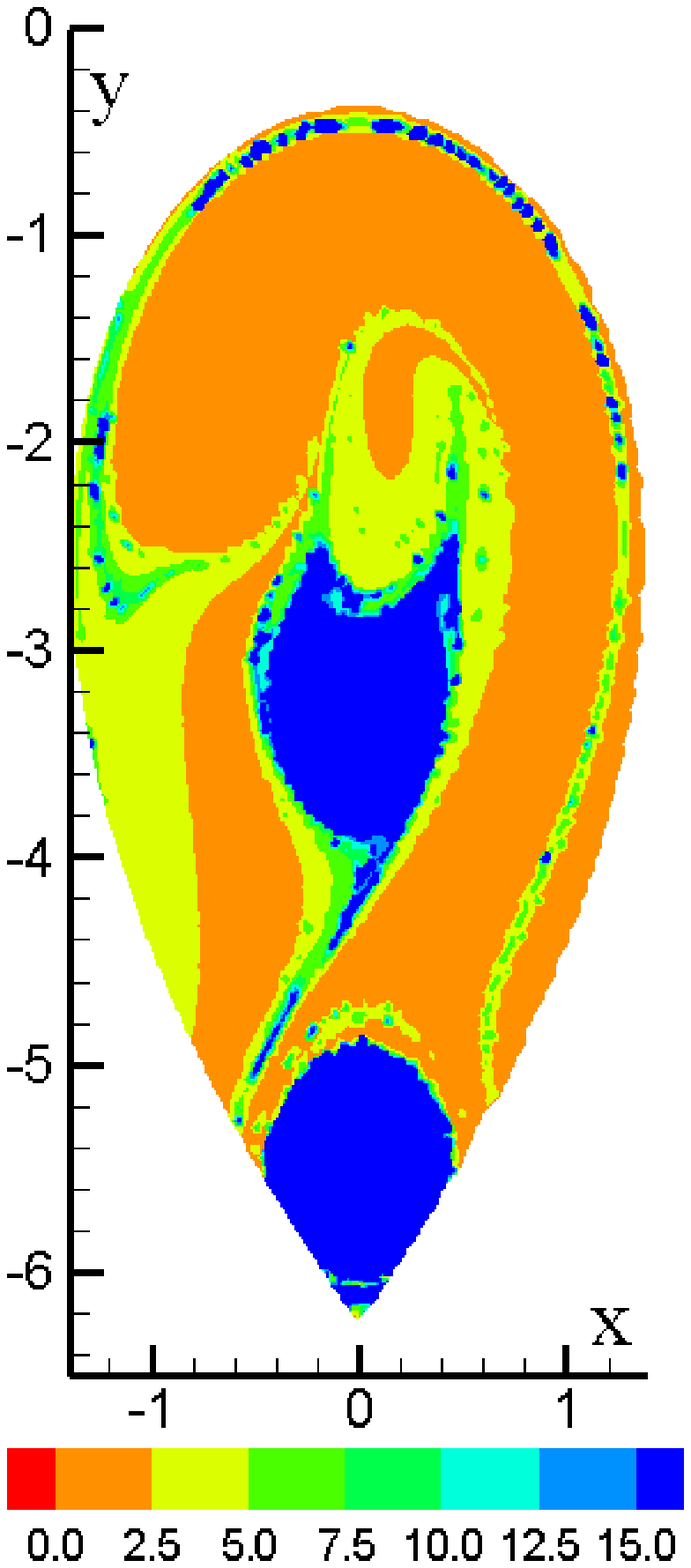}
\caption{$\kappa=0.1$,\\ $y=-5.5$} 
\end{subfigure}
\quad
\begin{subfigure}[t]{0.14\textwidth}
\includegraphics[width=2.5cm]{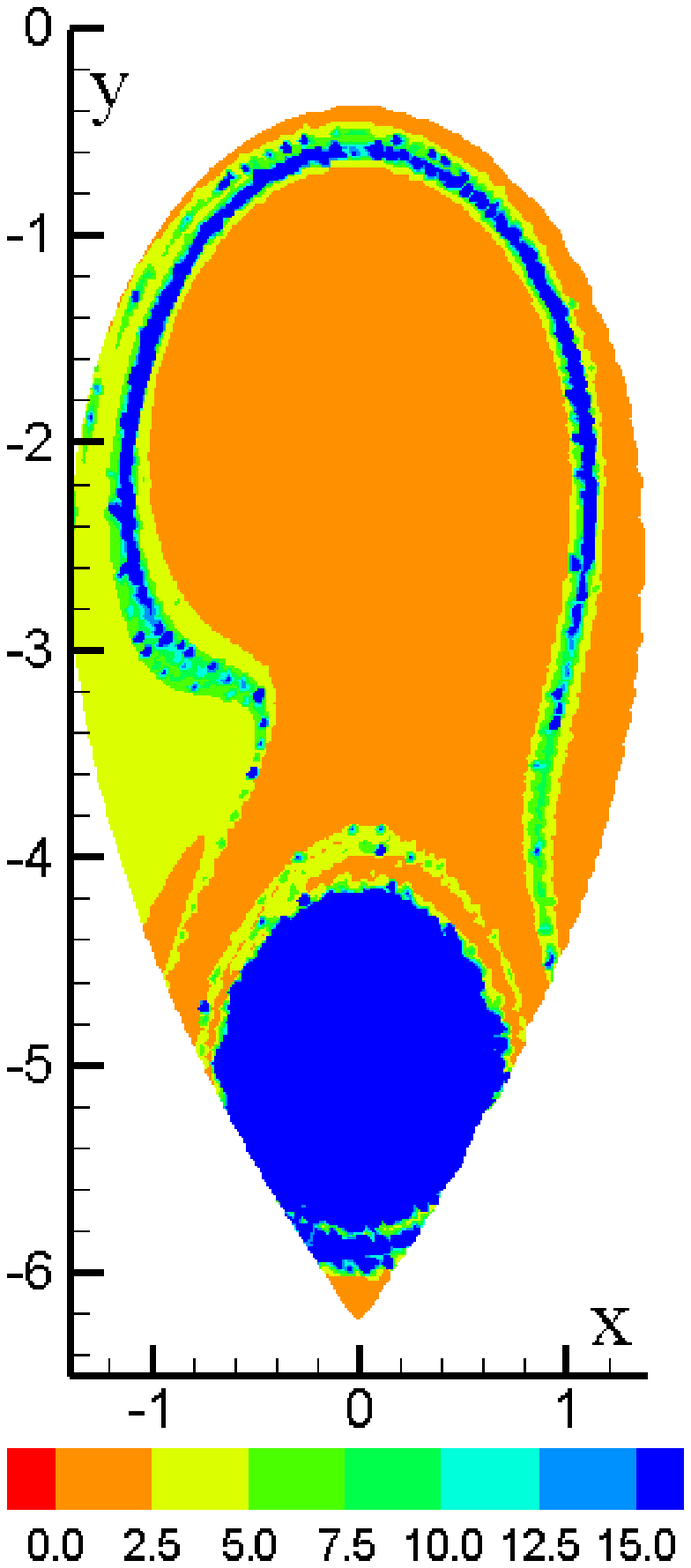}
\caption{$\kappa=0.3$,\\ $y=-5$} 
\end{subfigure}
\quad
\begin{subfigure}[t]{0.14\textwidth}
\includegraphics[width=2.5cm]{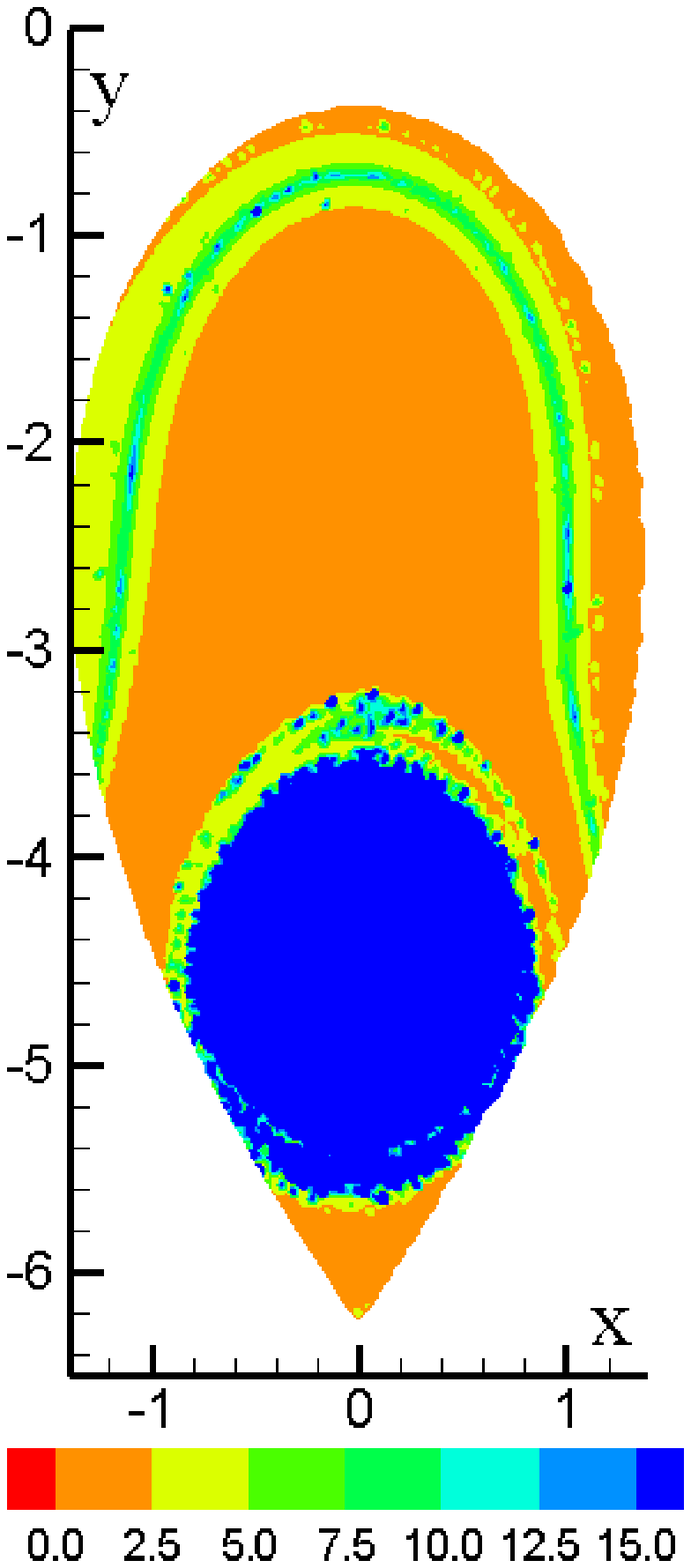}
\caption{$\kappa=0.5$,\\ $y=-4.5$} 
\end{subfigure}
\end{center}
\caption{Escaping time distribution in the upper-layer monopole propagation case.}
\label{fig5}
\end{figure*}

Also, as a Lagrangian advection measure, we have calculated the escaping time \citep{Kozlov_Koshel_1999, Kozlov_Koshel_2000, Izrailsky_et_al_2004}, which is determined as the time a fluid particle needs to be carried away by the exterior flow from the unperturbed topographic vortex region. This measure is an analogue for the Lyapunov exponent and it shows where Lagrangian advection progresses faster or slower. Thus, we have uniformly distributed within the separatrix $10^4$ markers, and, then, taken into consideration the time they would need to cross the line far enough out of the vortex interaction (line $x=5$). The escaping time distributions are shown in fig. \ref{fig5}, where unity of the time is equal to the corresponding period of a monopole revolution. A general feature of all the subfigures is the almost circle areas of long-live fluid particles. These areas correspond to the monopole region, which is very intense due to the singularity. Hence, fluid particles within these areas move mostly regular \citep{Ryzhov_Koshel_2011b}, and, therefore, they do not leave the topographic vortex region.

The blue color region corresponds to a strong influence of the monopole motion. At the initial stage of monopole motion, there are three regular critical points to form a heteroclinic structure (see fig. \ref{fig4}a). The topographic vortex cannot be distinctly identified due to no hyperbolic point corresponds to the unperturbed hyperbolic point. So, this case of monopole-topography interaction cannot be considered as a perturbation of the topographic vortex. Moreover, this initial stream-line picture resembles a counter-rotating dipole structure \citep[e.g.,][]{Voropayev_et_al_2001, Ryzhov_2011}. This structure changing in time results in that, at the half-period stage, there are also three regular critical points to form two homoclinic structures each associated either with the topographic or with the monopole vortices (see fig. \ref{fig4}b). Due to that topological alteration, fluid particle advection is very effective. Most particles are carried away within $5$ monopole revolutions (see fig. \ref{fig5}a).

The purple color region corresponds to a moderate influence of the monopole motion. At the initial stage, there are five regular critical points to form both one heteroclinic and one homoclinic structures (see fig. \ref{fig4}c). This homoclinic structure almost coincides with the unperturbed topographic vortex separatrix, that indicates this case can be considered as a perturbation of the topographic vortex. At the half-period stage, the stream-line picture appears as almost the same as in the aforementioned one of the blue color region (see fig. \ref{fig4}d). Lagrangian advection progresses although less efficient, but still very fast and it extends on the whole separatrix region. No stagnation zone appears within the region (see fig. \ref{fig5}b).

The grey color region corresponds to the least monopole influence. This positive $\kappa$ region differs topologically from those presented. Due to the vortices are co-rotating, the initial topological structure appears as a co-rotating dipole enveloped by a common topographic vortex separatrix (see fig. \ref{fig4}e). So, this structure can be considered as a topographic vortex with a double center. Although this double center greatly perturbs the fluid particle dynamics, the structure of the topographic vortex can be revealed during a whole monopole revolution (see fig. \ref{fig4}f). Due to the existence of two always unbroken centers, fluid particles in the vicinity of the topographic vortex center move almost regular, however, the surrounding fluid is carried away very fast (see fig. \ref{fig5}c).

The red color region corresponds to a transitional case of the monopole-topography interaction. Initially, the stream-line picture appears as a co-rotating dipole structure (see fig. \ref{fig4}g), although, during a monopole revolution, the dipole structure breaks, so, the singular monopole absorbs the topographic vortex elliptic point and becomes a new center of the topographic vortex for a certain time. During this time span, Lagrangian advection within the topographic vortex with the new singular center is rather regular (see fig. \ref{fig4}h). However, during a whole monopole revolution, almost all the fluid from the topographic vortex is carried away (see fig. \ref{fig5}d).

The green color region corresponds to the capturing of the monopole to be as a topographically trapped vortex with the singular monopole's center playing the role of a new topographic vortex center. Both at the initial and half-period stages, the stream-line portraits comprise only one regular critical point that corresponds to hyperbolic point of the topographic vortex. The initial stream-line portrait is shown in fig. \ref{fig4}i, while the half-period stream-line portrait appears as almost the same as that shown in fig. \ref{fig4}j. Thus, this case can be thought of as the topographic trapping of a monopole vortex. Lagrangian advection, in this case, differs insignificantly from the case previously addressed (see fig. \ref{fig5}e).

Now, we consider the diagram shown in fig. \ref{fig3}b for the middle-layer monopole propagation case. The main difference from the upper-layer monopole propagation case is, in this case, no singular points appear within the upper-layer velocity field, so, the monopole vortex appears as a regular one and as the regular topographic vortex also can be broken. Hence, a merger of the vortices can appear due to both vortices being regular. This regularity leads to all the half-period stream-line pictures to appear almost the same with one elliptic point, which is formed by the merger, and one hyperbolic point. Also, the lack of a singular point leads to that Lagrangian advection is much more regular in comparison with that considered above.

\begin{figure}[t]
\vspace*{2mm}
\begin{center}
\begin{subfigure}[t]{0.21\textwidth}
\includegraphics[width=3.5cm]{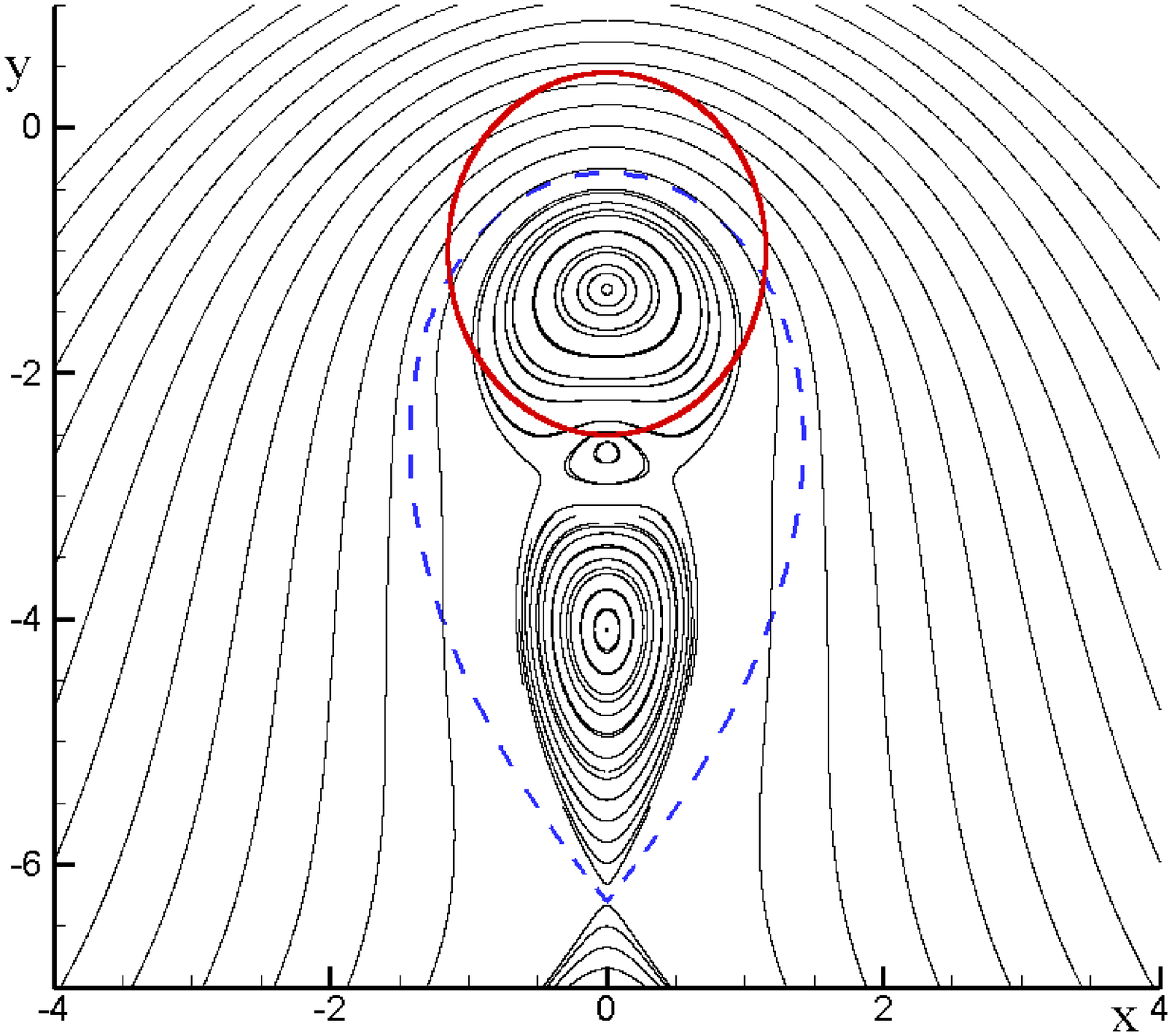}
\caption{initial stage for the yellow color region $\left(\kappa=-0.7,y=-2.5\right)$} 
\end{subfigure}
\quad
\begin{subfigure}[t]{0.21\textwidth}
\includegraphics[width=3.5cm]{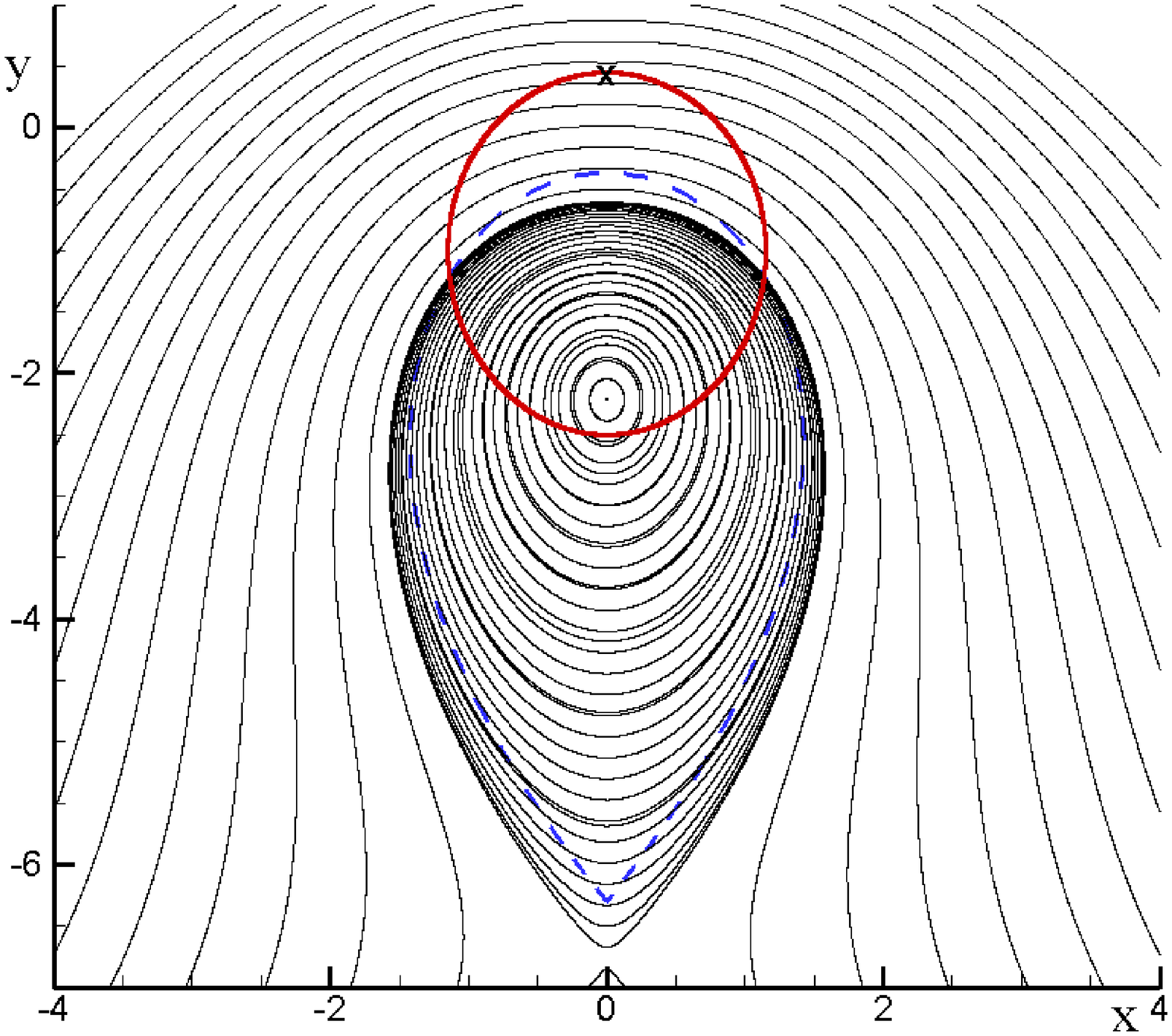}
\caption{half-period stage  for the yellow color region $\left(\kappa=-0.7,y=-2.5\right)$} 
\end{subfigure}
\begin{subfigure}[t]{0.21\textwidth}
\includegraphics[width=3.5cm]{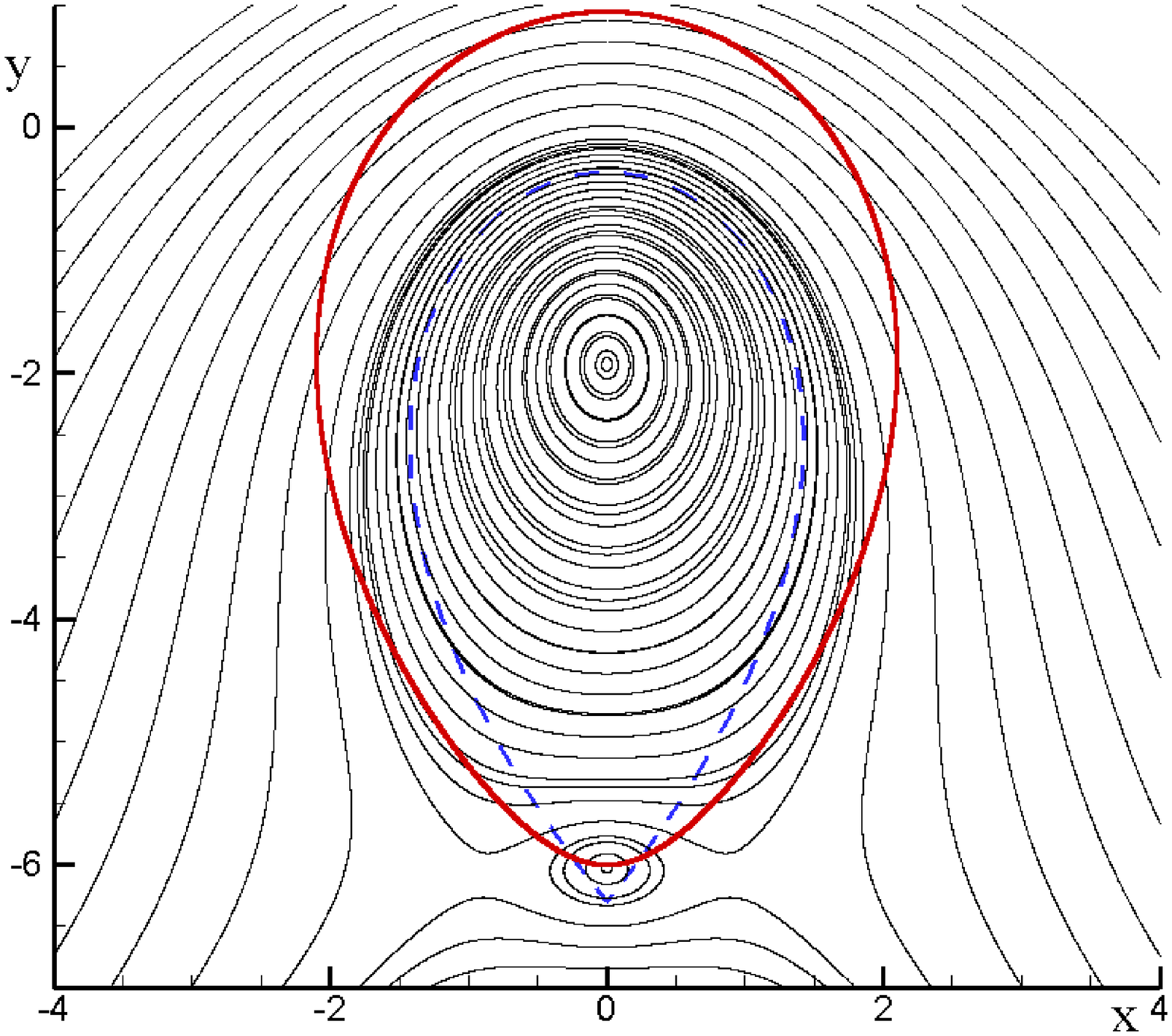}
\caption{initial stage for the negative orange color region $\left(\kappa=-0.5,y=-6\right)$} 
\end{subfigure}
\quad
\begin{subfigure}[t]{0.21\textwidth}
\includegraphics[width=3.5cm]{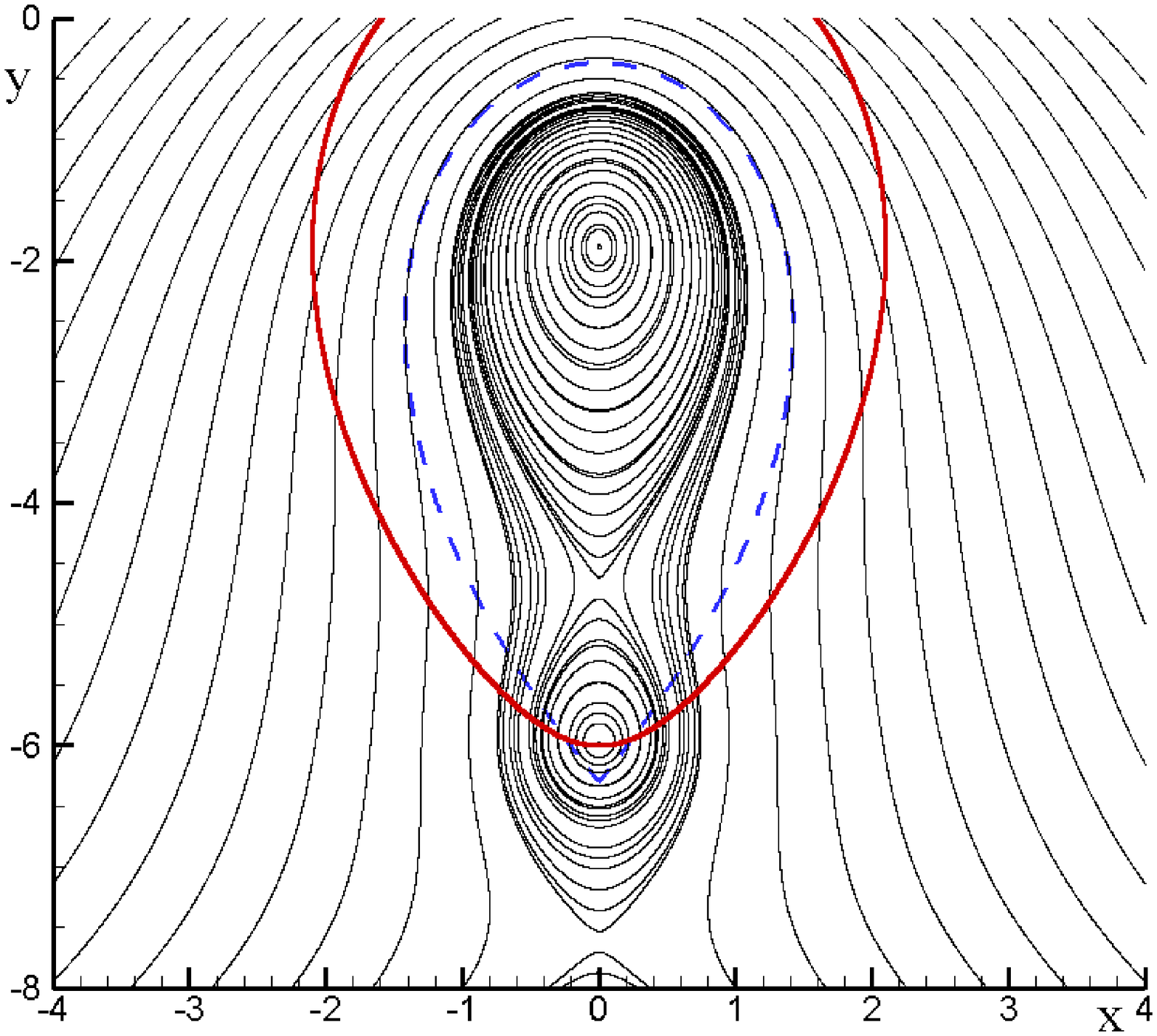}
\caption{initial stage for the positive orange color region $\left(\kappa=0.5,y=-6\right)$} 
\end{subfigure}
\end{center}
\caption{Stream-lines of the flow in the middle-layer monopole propagation case. Red curve corresponds to the monopole motion trajectory. Dashed blue curve is the topographic vortex unperturbed separatrix.}
\label{fig6}
\end{figure}

\begin{figure*}[t]
\vspace*{2mm}
\begin{center}
\begin{subfigure}[t]{0.14\textwidth}
\includegraphics[width=2.5cm]{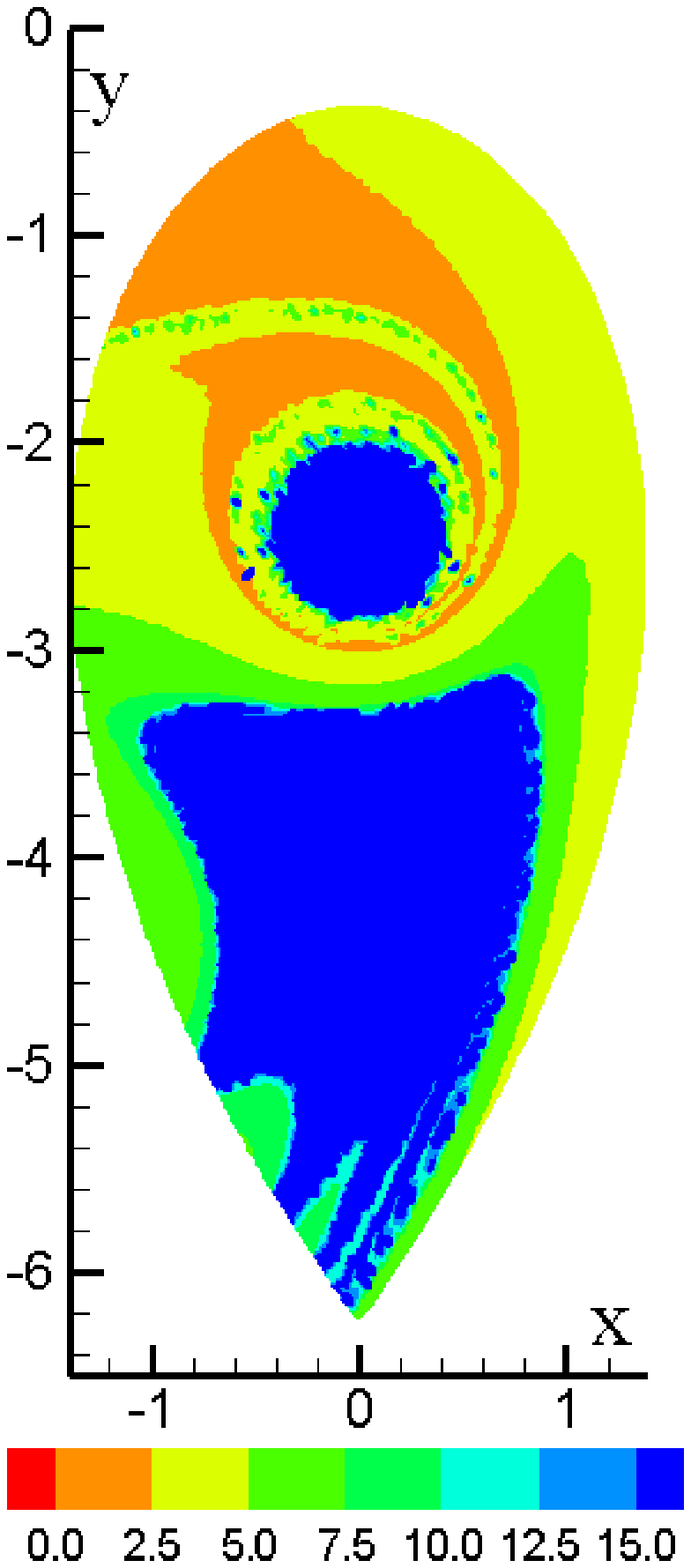}
\caption{$\kappa=-0.7$,\\ $y=-2.5$} 
\end{subfigure}
\quad
\begin{subfigure}[t]{0.14\textwidth}
\includegraphics[width=2.5cm]{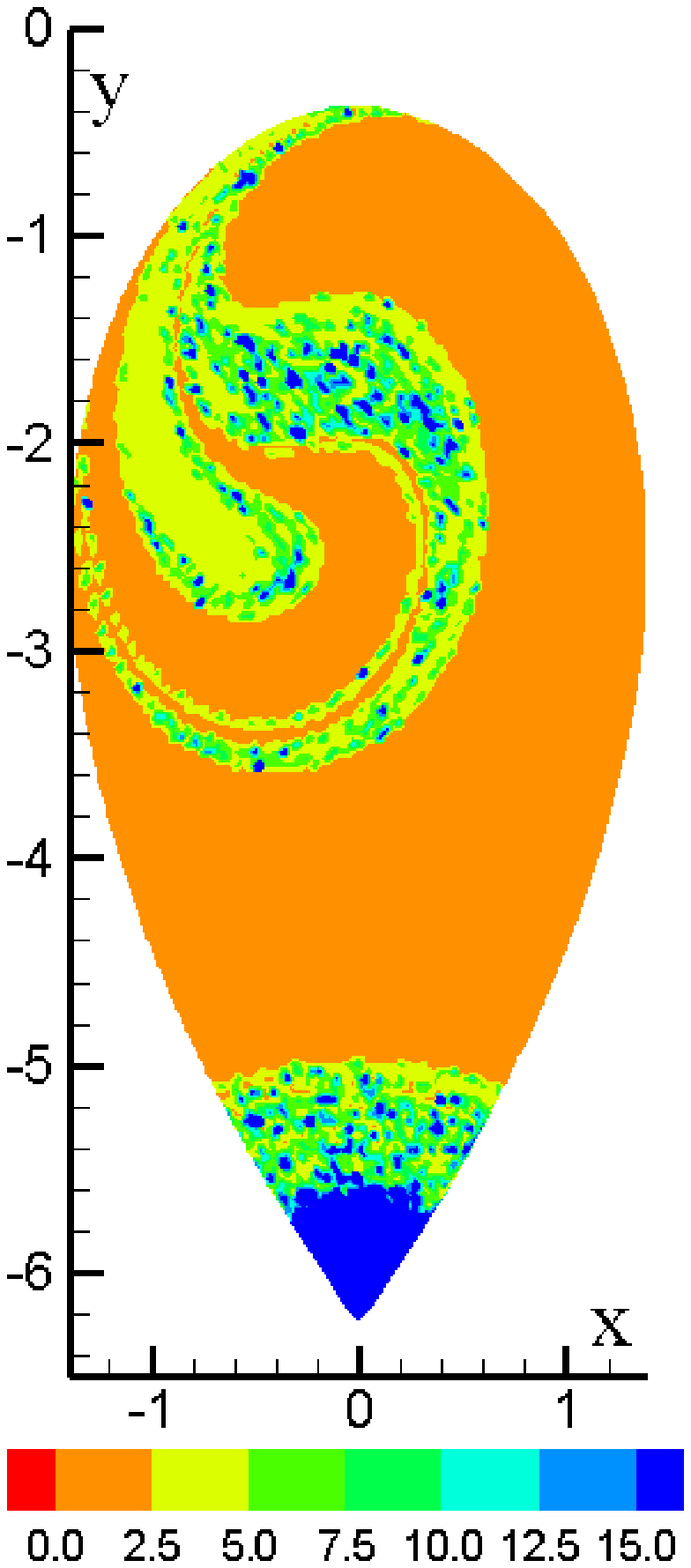}
\caption{$\kappa=-0.5$,\\ $y=-6$} 
\end{subfigure}
\quad
\begin{subfigure}[t]{0.14\textwidth}
\includegraphics[width=2.5cm]{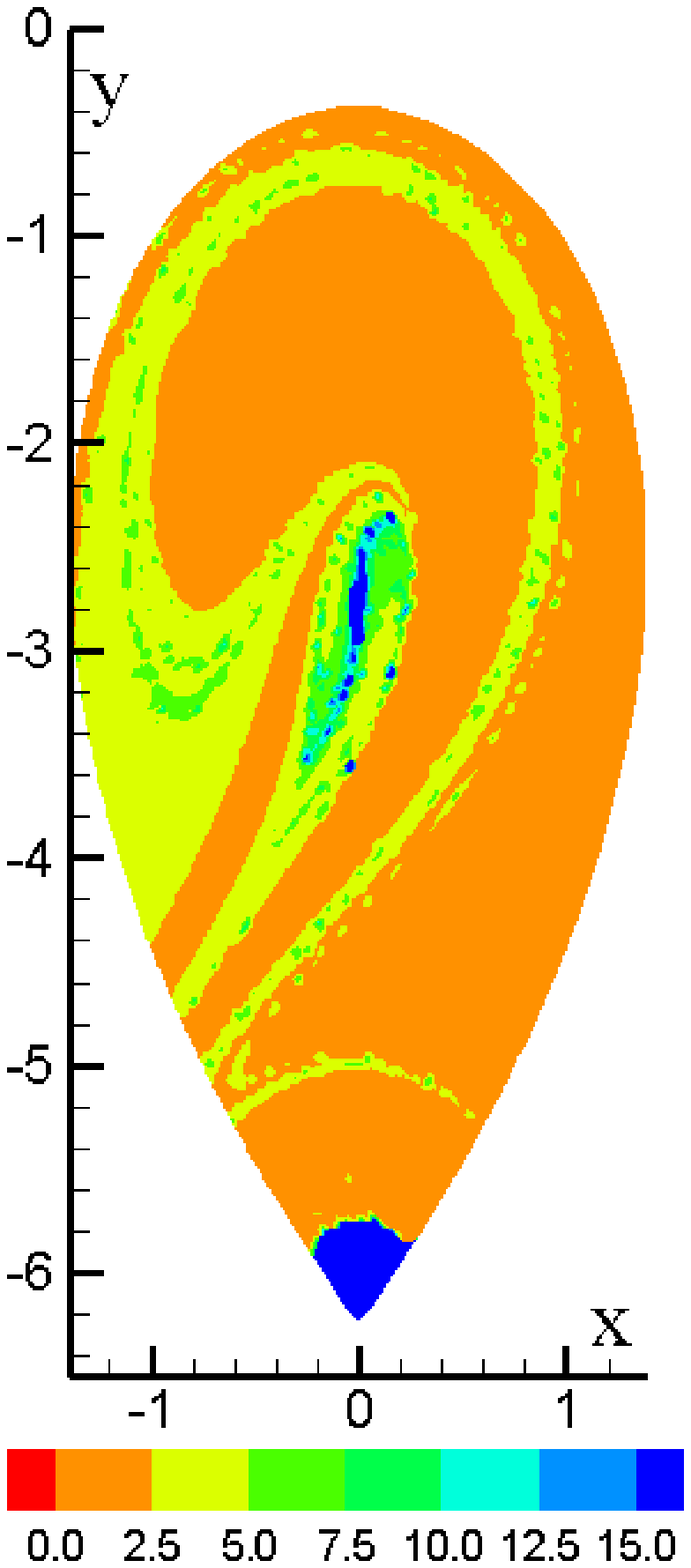}
\caption{$\kappa=0.5$,\\ $y=-6$} 
\end{subfigure}
\quad
\begin{subfigure}[t]{0.14\textwidth}
\includegraphics[width=2.5cm]{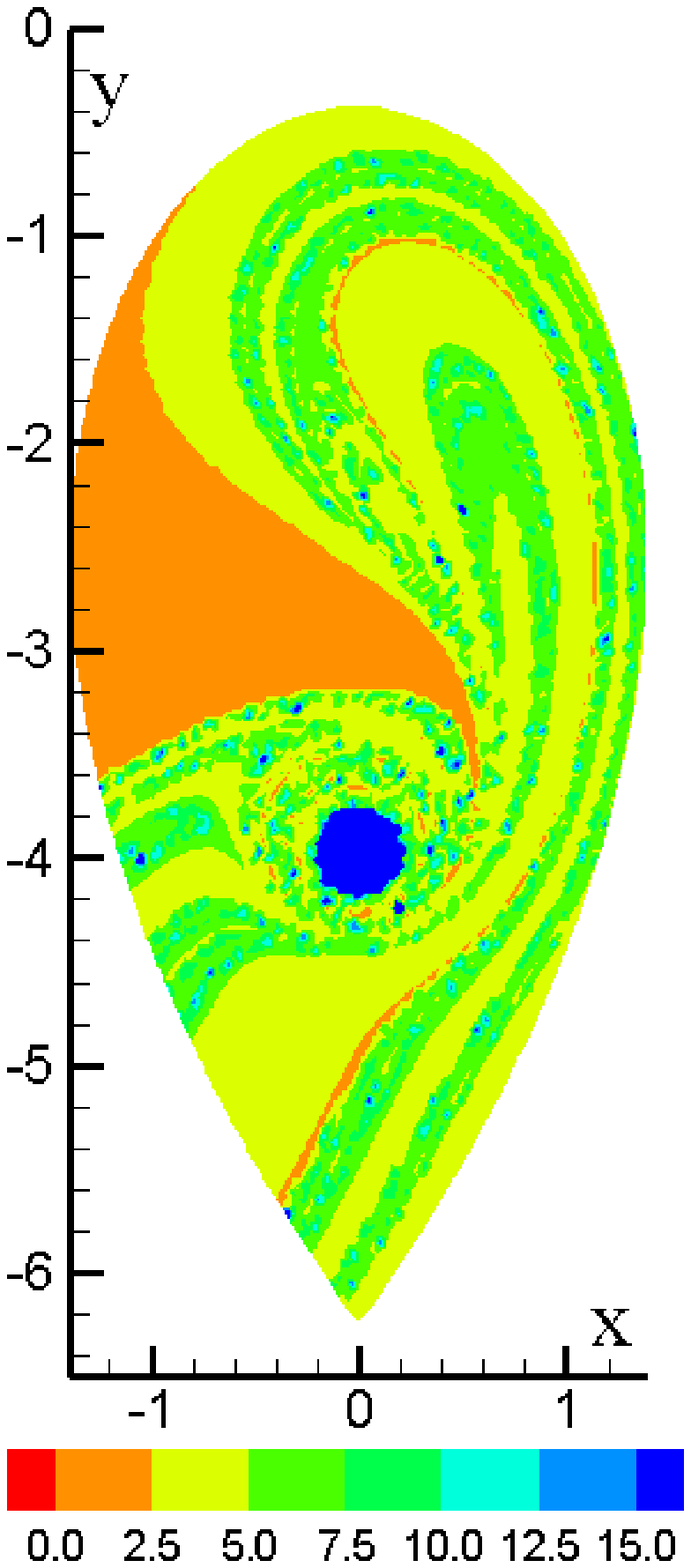}
\caption{$\kappa=-0.2$,\\ $y=-4$} 
\end{subfigure}
\quad
\begin{subfigure}[t]{0.14\textwidth}
\includegraphics[width=2.5cm]{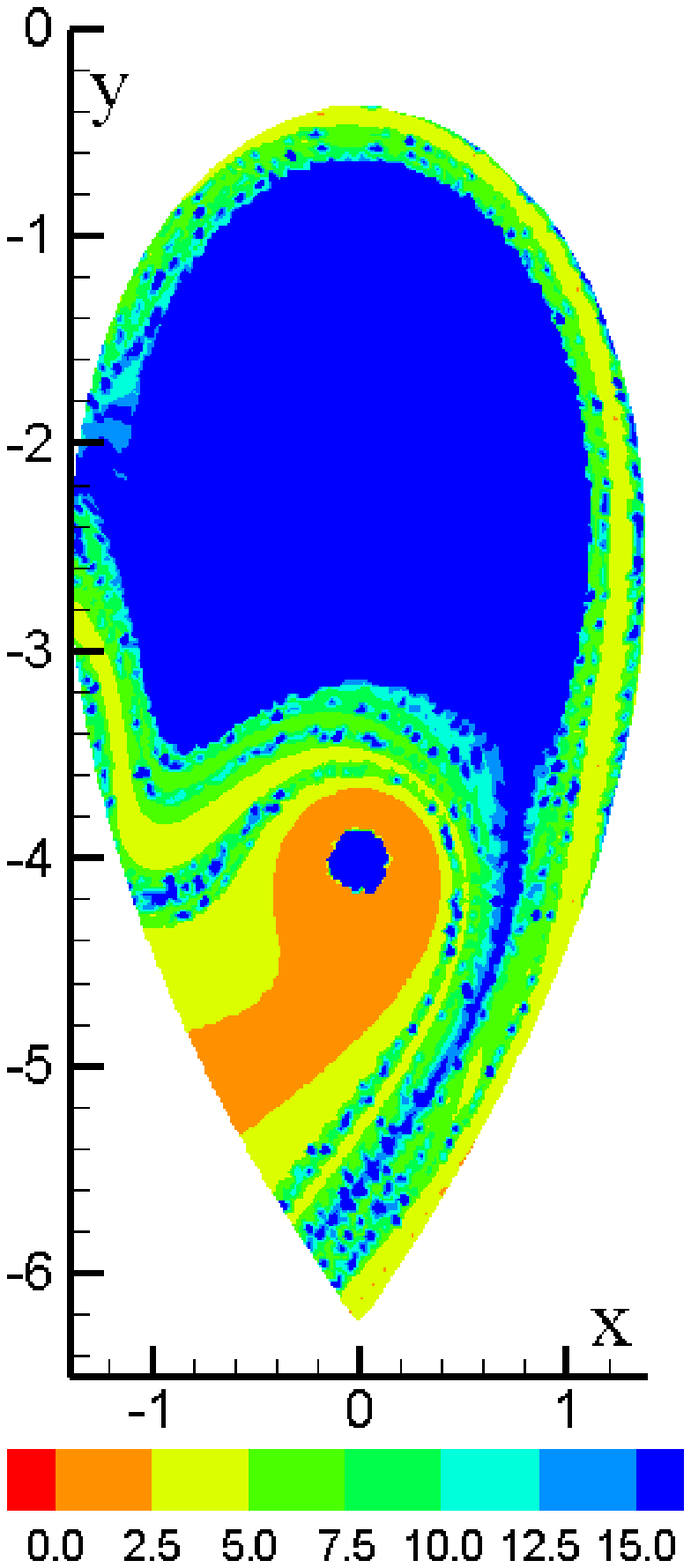}
\caption{$\kappa=0.2$,\\ $y=-4$} 
\end{subfigure}
\end{center}
\caption{Escaping time distribution in the middle-layer monopole propagation case.}
\label{fig7}
\end{figure*}

The yellow color region corresponds to six initial regular critical points (see fig. \ref{fig6}a) and two half-period critical points (see fig. \ref{fig6}b), that both correspond to the topographic vortex with the monopole vortex being disappeared since the velocity of surrounding flow is too high for a closed circulation region to be formed. The position of the corresponding middle-layer monopole vortex is marked by the cross. Such a half-period stream-line portrait is universal for all the color regions shown in fig. \ref{fig6}b. The corresponding escaping time distribution is shown in fig. \ref{fig7}a. There is a big stagnation region with mostly regular advection corresponding to the lower closed region shown in fig. \ref{fig6}a.

The orange color region is arranged astride the $\kappa=0$ line. This region corresponds to the existence of three initial critical points. The difference between the negative and positive orange color region initial stream-line portraits is shown in fig. \ref{fig6}c,d. Both the corresponding half-period stream-line portraits, however, appear as almost the same as that shown in fig. \ref{fig6}b. Since initially two vortex structures can be reliably identified, and at half-period stage all these structures merge, the escaping time distribution shows very effective and intense advection with no stagnation regions progressing. Figures \ref{fig7}c,d depict the escaping time distribution in the negative and positive $\kappa$ cases, respectively.

The brown color region corresponds to the existence of two initial and half-period critical points. The middle-layer monopole does not induce a closed region within the upper layer. Despite that, the middle-layer monopole does greatly perturb fluid particle advection. The corresponding stream-line portrait does not change topologically during a monopole revolution and it appears as almost the same as that shown in fig. \ref{fig6}b. However, on both sides of the $\kappa=0$ line, the advection efficiency is very different. In the $\kappa<0$ zone, advection is very irregular (see fig. \ref{fig7}d) due to counter-rotation of the middle-layer monopole and the topographic vortex. On the other hand, in the $\kappa>0$ zone, advection is mostly regular, a big stagnation region appears in region of the topographic vortex (see fig. \ref{fig7}e), due to co-rotation of the middle-layer monopole and the topographic vortex.

Further, we study how irregular motion of the monopole to influence Lagrangian advection
\subsection{Irregular monopole motion}
In this paragraph, we analyze fluid particle advection being induced by a non-periodic perturbation consisting of periodic background flow oscillation and non-periodic part due to monopole irregular motion within the topographic vortex. That irregular monopole motion is due to the monopole's singular center moving as a fluid particle in the periodically driven velocity field of the topographic vortex, which is known to produce the irregular dynamics   \citep[e.g.,][]{Sokolovskiy_et_al_1998, Kozlov_Koshel_2001, Izrailsky_et_al_2004, Izrailsky_et_al_2008, Koshel_et_al_2008}. Strictly speaking, if one set the background flow to oscillate periodically,

\begin{equation}
\label{eq15}
W = {W_0}\left( {1 + {\mu _W}\cos {\nu _W}t} \right),
\end{equation}
where $\mu _W$, and $\nu _W$ are the magnitude and frequency of the background flow oscillation, then system (\ref{eq13}) becomes a system with one and a half degree of freedom permitting the chaotic dynamics to occur. Hence, with such an oscillating background flow, the monopole can start moving out of the topographic vortex, and then it can be trapped temporarily by the topography. And, on the contrary, if the monopole starts moving within the topographic vortex, it now can be carried away by the exterior flow. It should be mentioned that the background flow oscillation also affects the fluid particle dynamics, resulting in certain number of particles to leave the topographic vortex region. However, in our numerical simulation, we chose a very small perturbation magnitude ($\mu_W=0.01$), so there are very few such particles. So, by making use of such a configuration, we study Lagrangian advection being mostly induced by the short-term monopole-topography interaction.

\begin{figure*}[t]
\vspace*{2mm}
\begin{center}
\begin{subfigure}[t]{0.14\textwidth}
\includegraphics[width=2.5cm]{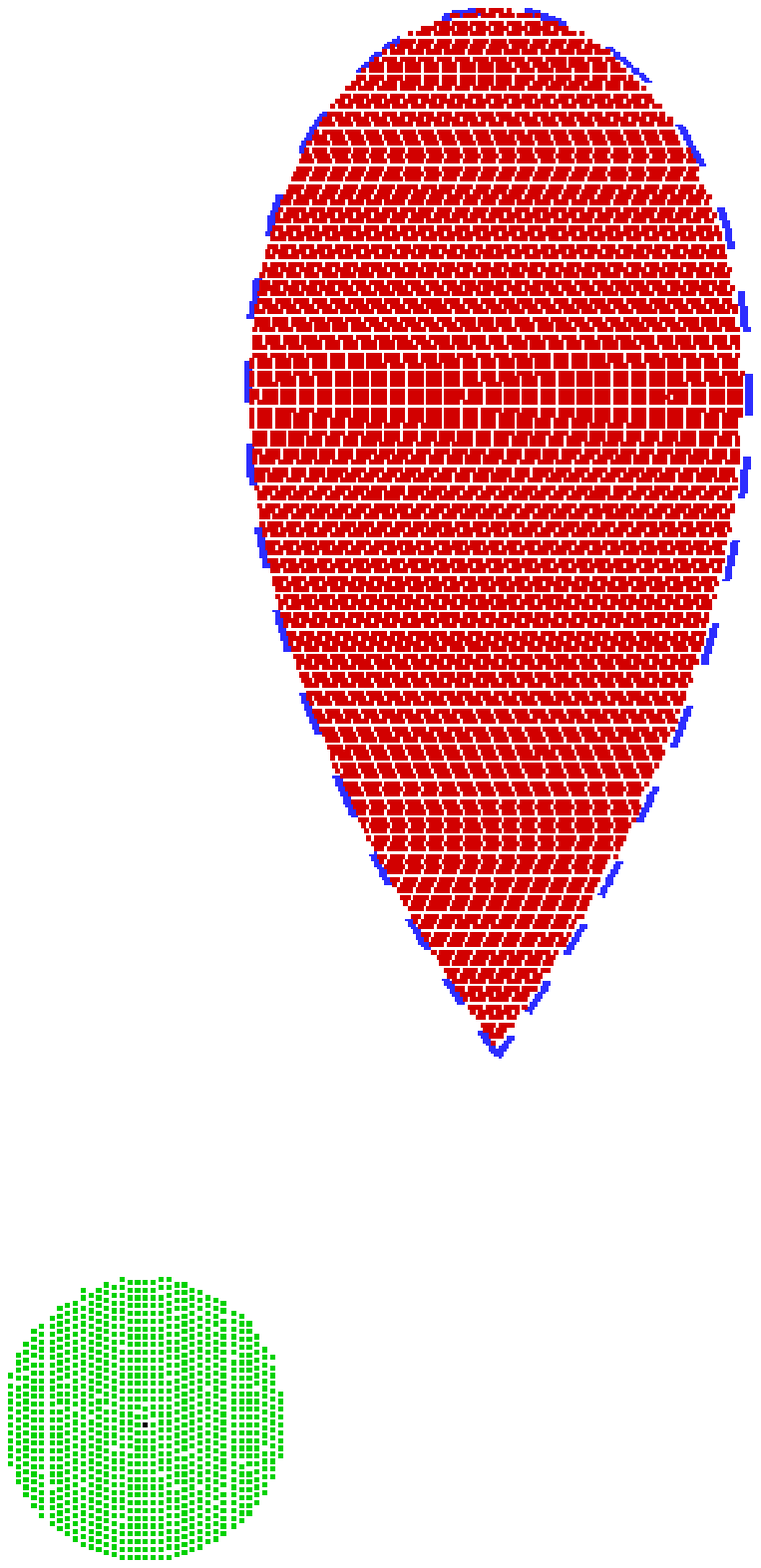}
\caption{$t=0$} 
\end{subfigure}
\quad
\begin{subfigure}[t]{0.14\textwidth}
\includegraphics[width=2.5cm]{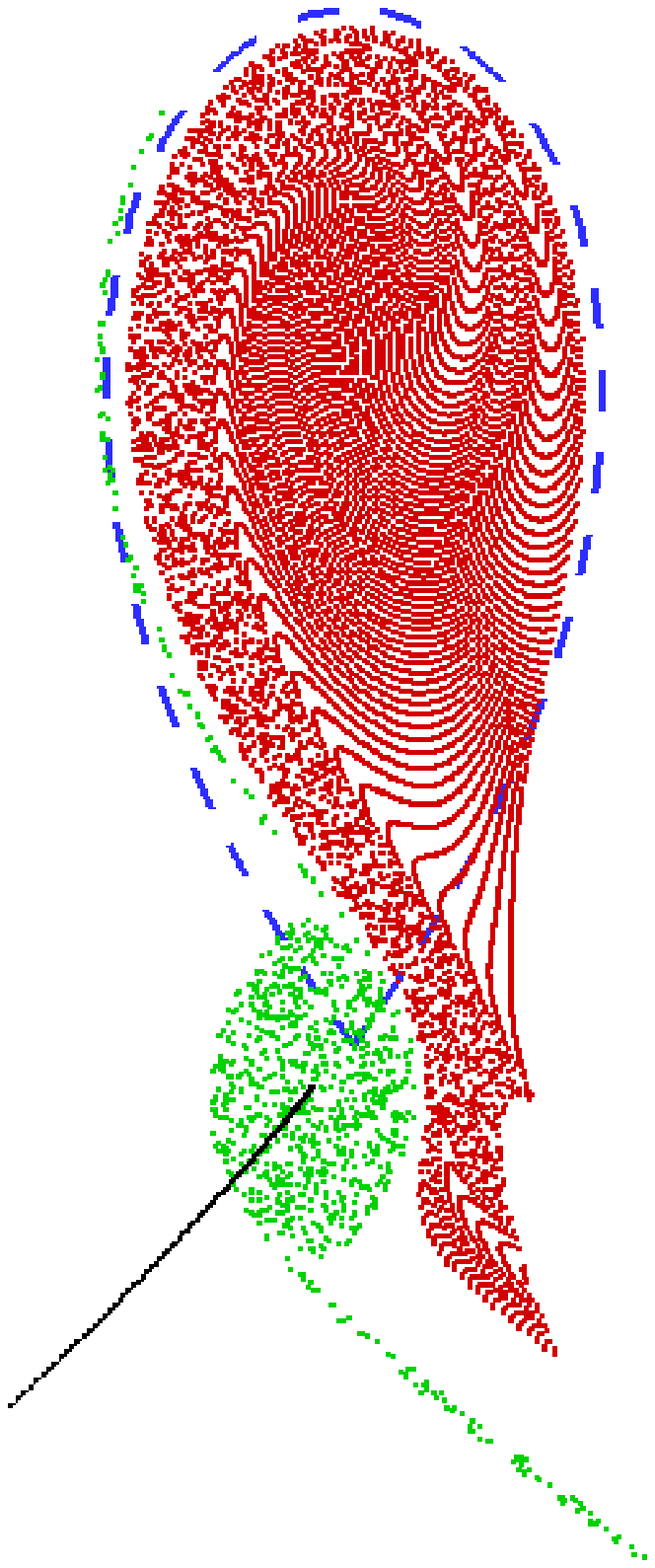}
\caption{$t=30$} 
\end{subfigure}
\quad
\begin{subfigure}[t]{0.14\textwidth}
\includegraphics[width=2.5cm]{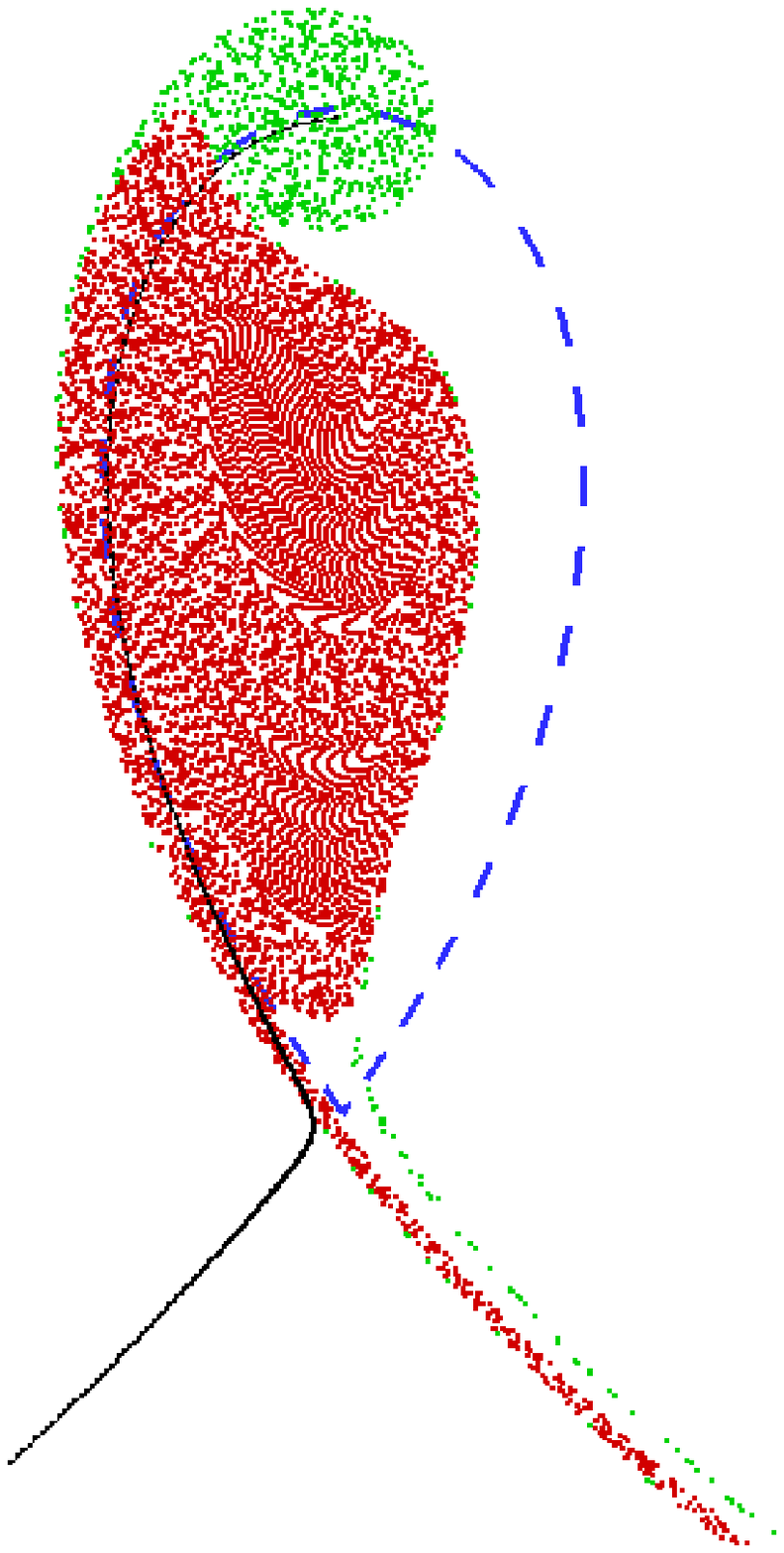}
\caption{$t=90$} 
\end{subfigure}
\quad
\begin{subfigure}[t]{0.14\textwidth}
\includegraphics[width=2.5cm]{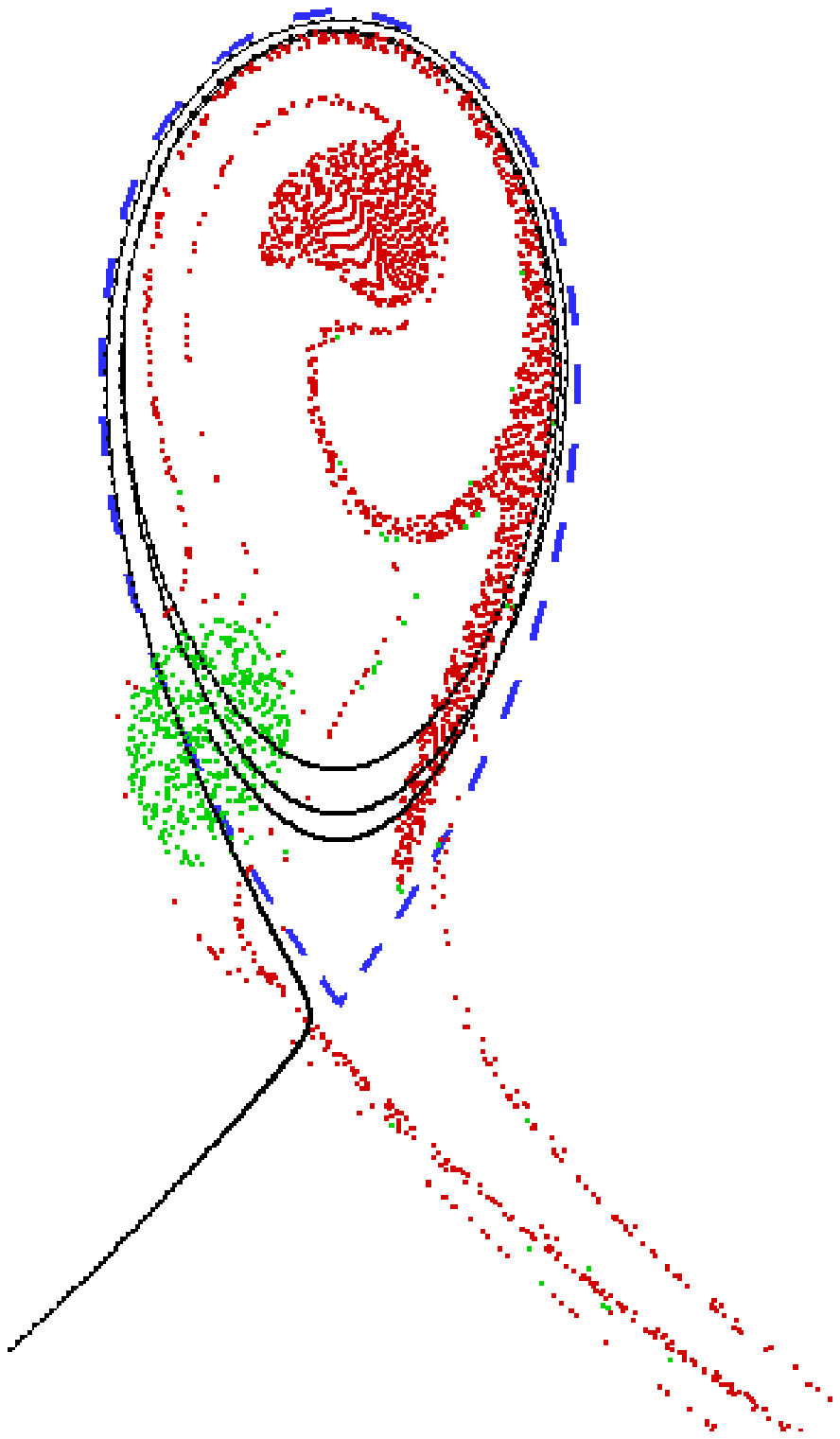}
\caption{$t=240$} 
\end{subfigure}
\quad
\begin{subfigure}[t]{0.14\textwidth}
\includegraphics[width=2.5cm]{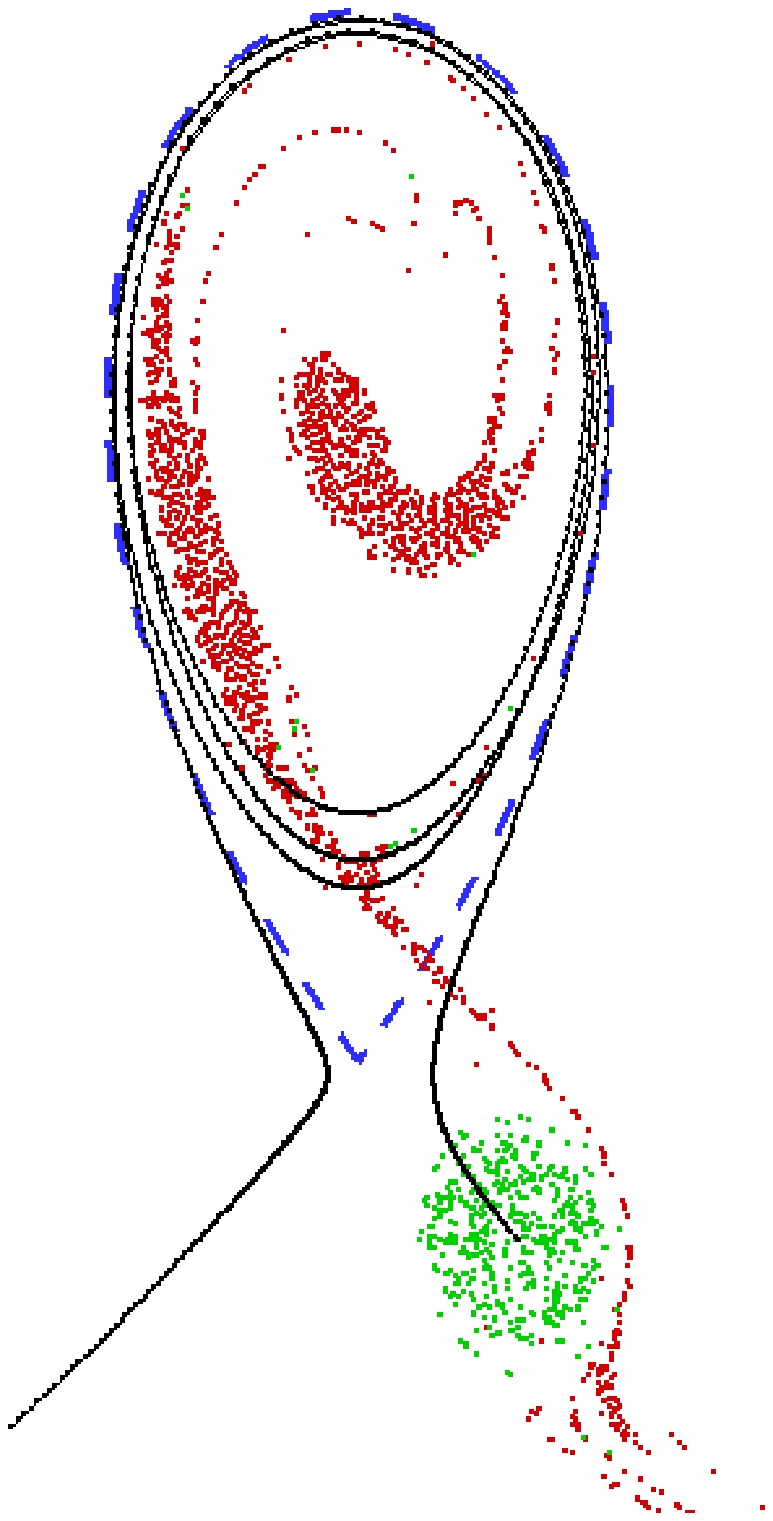}
\caption{$t=315$} 
\end{subfigure}
\end{center}
\caption{Upper-layer monopole propagation case. Particle scattering at the short-term monopole-topography interaction. Red and green markers correspond to the topographic and monopole vortex regions, respectively, the blue dashed curve is the unperturbed topographic vortex separatrix, and the black curve points out the trajectory of the monopole's center. Subfigures depict markers distribution at the corresponding instant in time.}
\label{fig8}
\end{figure*}

Figure \ref{fig8} depicts an example of fluid particle advection being generated by the short-term interaction, while the monopole accomplishes a few revolutions within the topographic vortex. Figure \ref{fig8}a shows the initial configuration of red and green markers corresponding to the topographic and monopole vortex regions, respectively. The unperturbed topographic vortex region is uniformly filled in with $10^4$ red markers. Also, $1.5\cdot 10^3$ green markers are placed to distinguish the monopole vortex region. The monopole with strength $\kappa=0.1$ starts moving out of the topographic vortex (see fig. \ref{fig8}a) at the position with coordinates $x=-2,\; y=-8.4$. Then, the monopole vortex is captured by the topographic vortex due to chaotic advection (see fig. \ref{fig8}b). Next fig. \ref{fig8}c shows the marker distribution as the monopole has passed a half of rotational period (the black curve points out the trajectory of the monopole's center). A great deformation caused by the monopole is clearly seen. Figure \ref{fig8}d illustrates the particle distribution after the monopole has made three whole revolutions about the topography. Few red markers from the initial distribution have stayed within the topographic region. Last fig. \ref{fig8}e depicts the monopole leaving the topographic vortex region after four revolutions.

\begin{figure*}[t]
\vspace*{2mm}
\begin{center}
\begin{subfigure}[t]{0.14\textwidth}
\includegraphics[width=2.5cm]{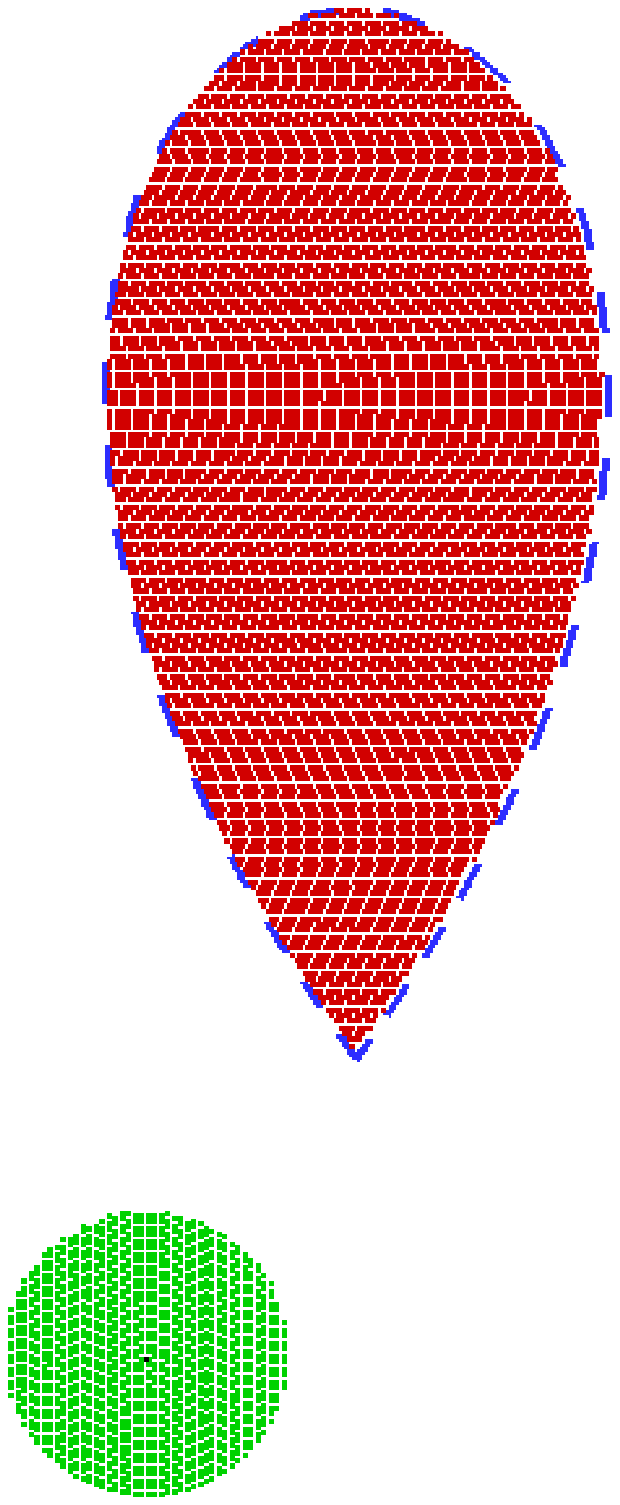}
\caption{$t=0$} 
\end{subfigure}
\quad
\begin{subfigure}[t]{0.14\textwidth}
\includegraphics[width=2.5cm]{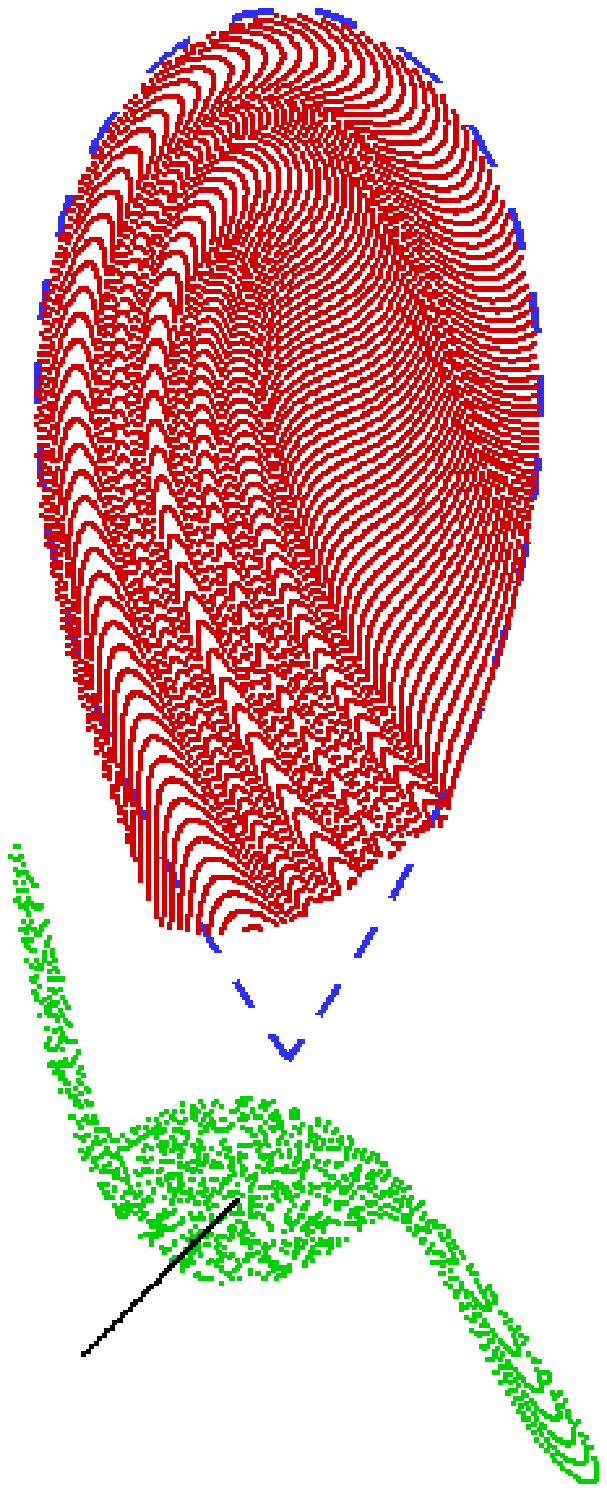}
\caption{$t=18$} 
\end{subfigure}
\quad
\begin{subfigure}[t]{0.14\textwidth}
\includegraphics[width=2.5cm]{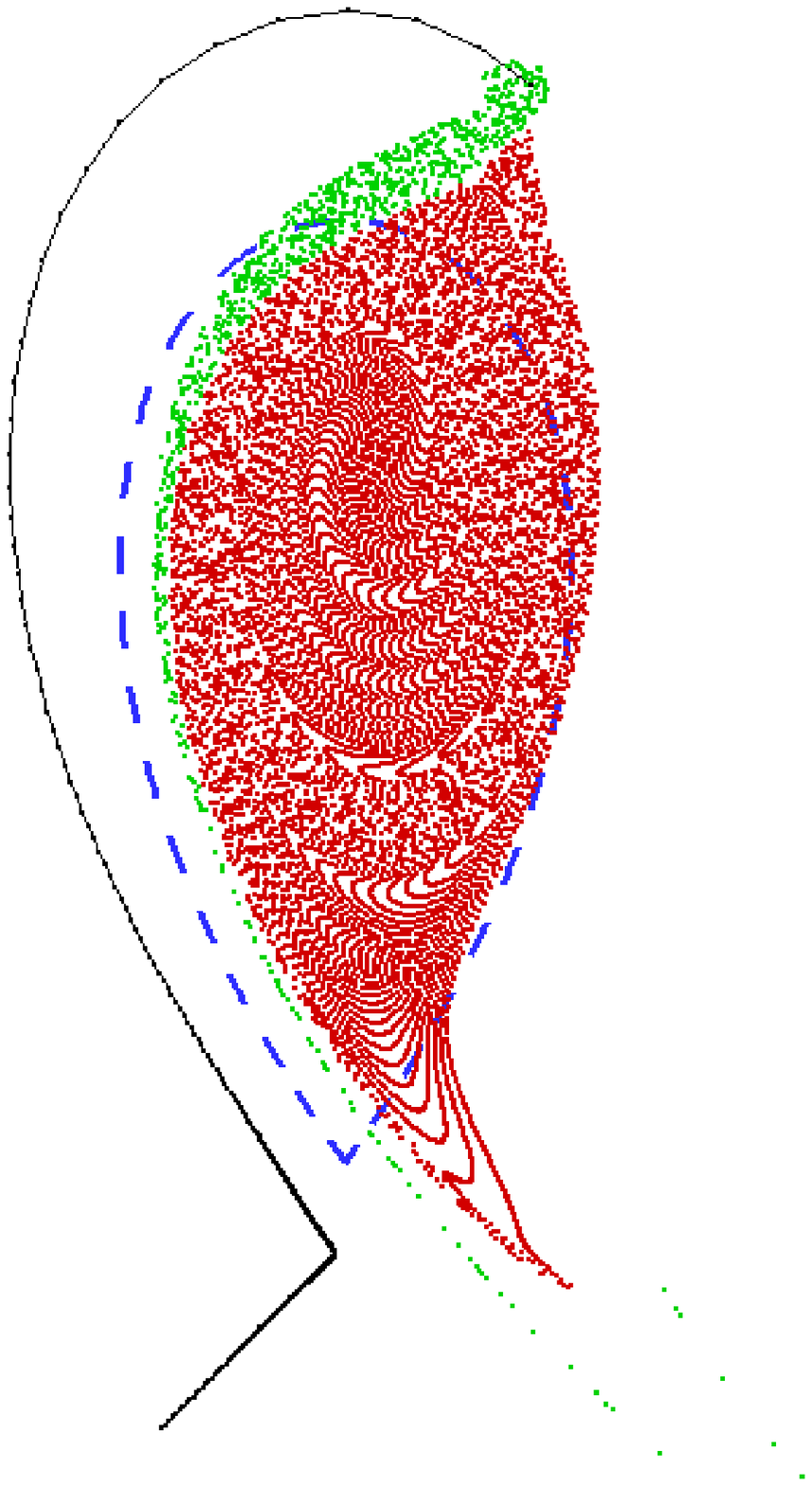}
\caption{$t=96$} 
\end{subfigure}
\quad
\begin{subfigure}[t]{0.14\textwidth}
\includegraphics[width=2.5cm]{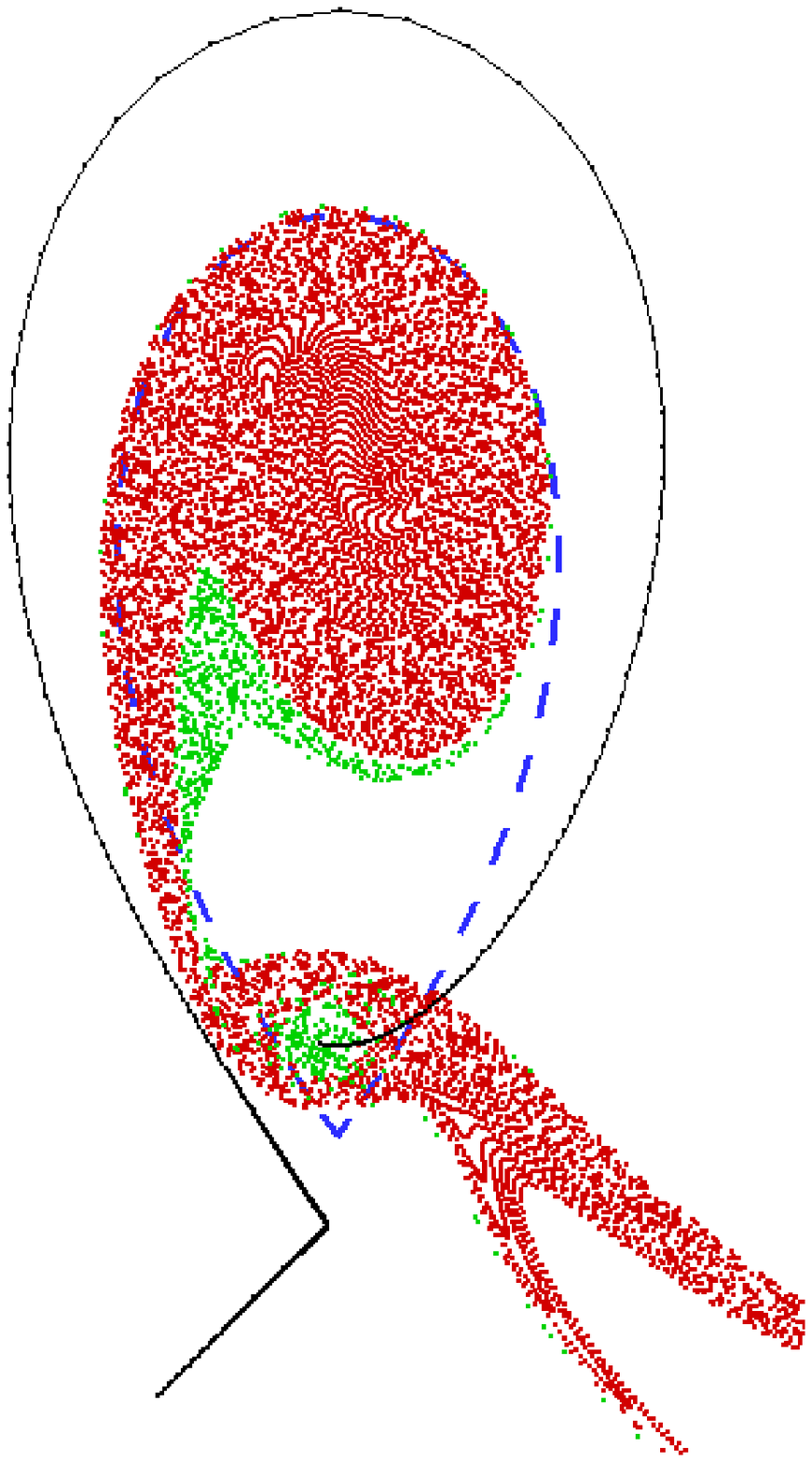}
\caption{$t=120$} 
\end{subfigure}
\quad
\begin{subfigure}[t]{0.14\textwidth}
\includegraphics[width=2.5cm]{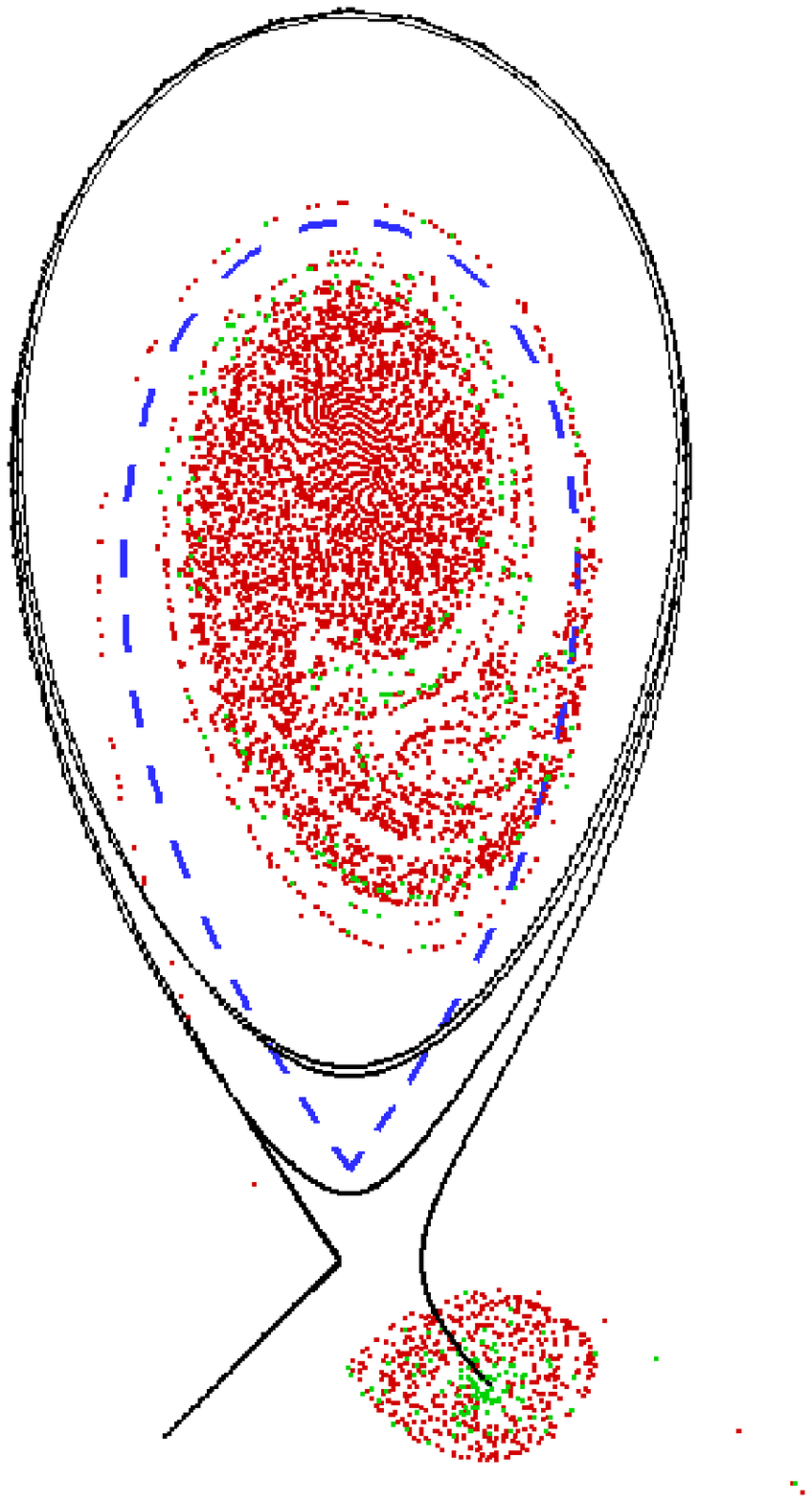}
\caption{$t=315$} 
\end{subfigure}
\end{center}
\caption{The same as in fig. \ref{fig8} for the middle-layer monopole propagation case.}
\label{fig8_}
\end{figure*}

Figure \ref{fig8_} also depicts a series of marker scattering patterns, but for the middle-layer monopole propagation case. In this case, the monopole starts moving at the position with coordinates $x=-1.18,\; y=-8$ and it appears as a regular vortex within the upper layer. It results in that a closed recirculation region corresponding to the monopole ceases to exist at the half-period stage. Hence, the green markers mostly leave the monopole region (see fig. \ref{fig8_}c). However, when the closed recirculation region appears again (see fig. \ref{fig8_}d), the monopole captures a great deal of the red markers initially associated with the topographic vortex. Thus, during the topography capturing, the monopole encloses some red markers, then after being carried away from the topographic vortex, it advects them to the infinity (see fig. \ref{fig8_}e).

Figures \ref{fig8} and \ref{fig8_} also clearly shows that the particle advection is greatly affected to the number of monopole revolutions about the topography. The longer the monopole revolves about the topography the more effective advection is. To estimate that short-term monopole influence, we have performed a numerical simulation, in which we calculate the number of fluid particles escaping the topographic vortex with respect to the number of the monopole revolutions. Since the monopole motion is irregular, two initially close monopole trajectories wind the topography very differently, with different revolution numbers. Hence, it is impossible to predict how many revolutions complete the monopole starting at a new initial position. Thus, as initial positions for the monopole, we have chosen two intervals of initial positions $\left(x=-2,\; y \in \left[-8.42;-8.38\right] \right)$ for the upper-layer monopole propagation case, and $\left(x=-1.18,\; y \in \left[-8.02;-7.98\right] \right)$ for the middle-layer monopole propagation case. 

\begin{figure*}[t]
\vspace*{2mm}
\begin{center}
\begin{subfigure}[t]{0.4\textwidth}
\includegraphics[width=5.3cm]{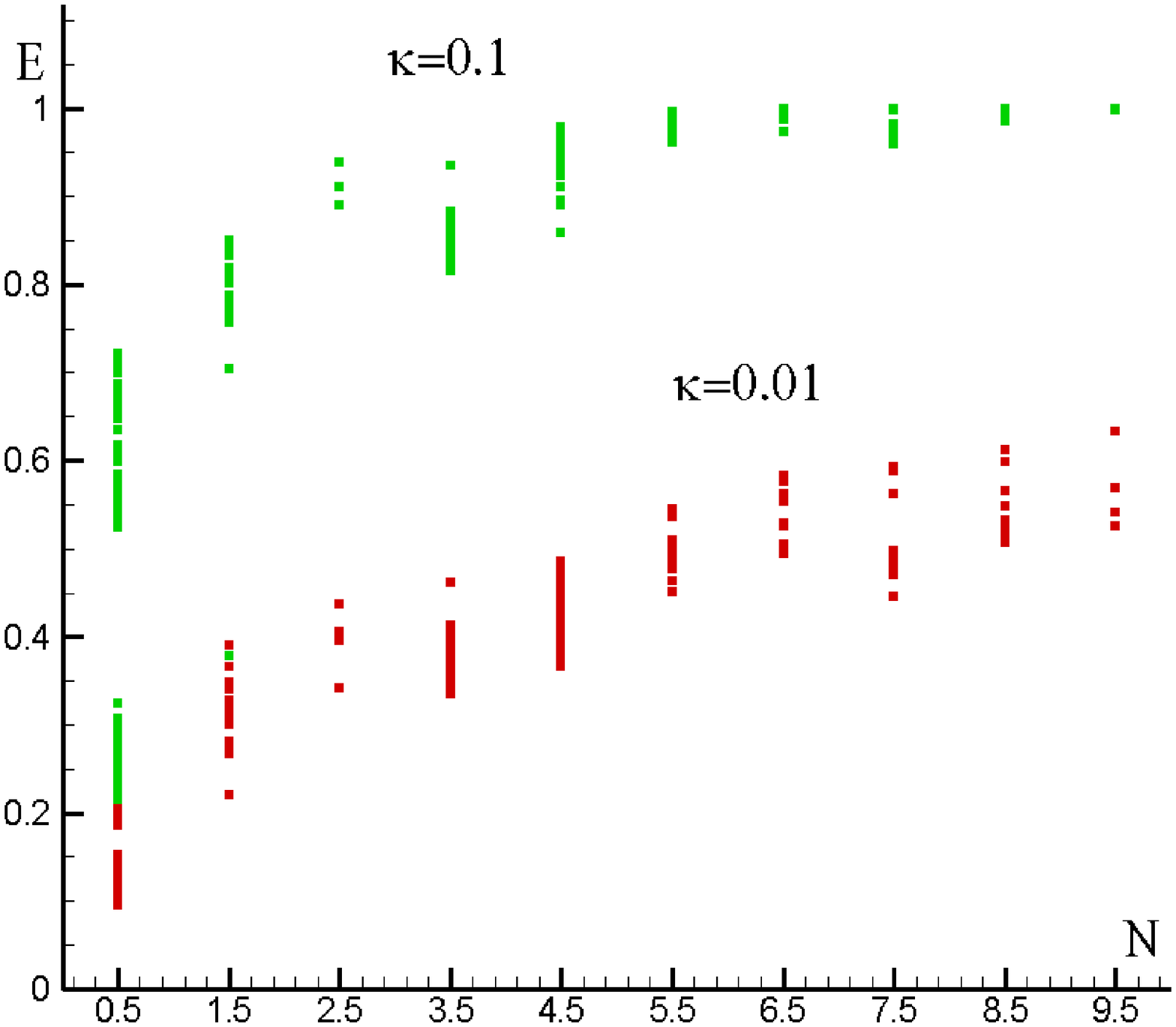}
\caption{upper-layer positive $\kappa$ monopole} 
\end{subfigure}
\quad
\begin{subfigure}[t]{0.4\textwidth}
\includegraphics[width=5.3cm]{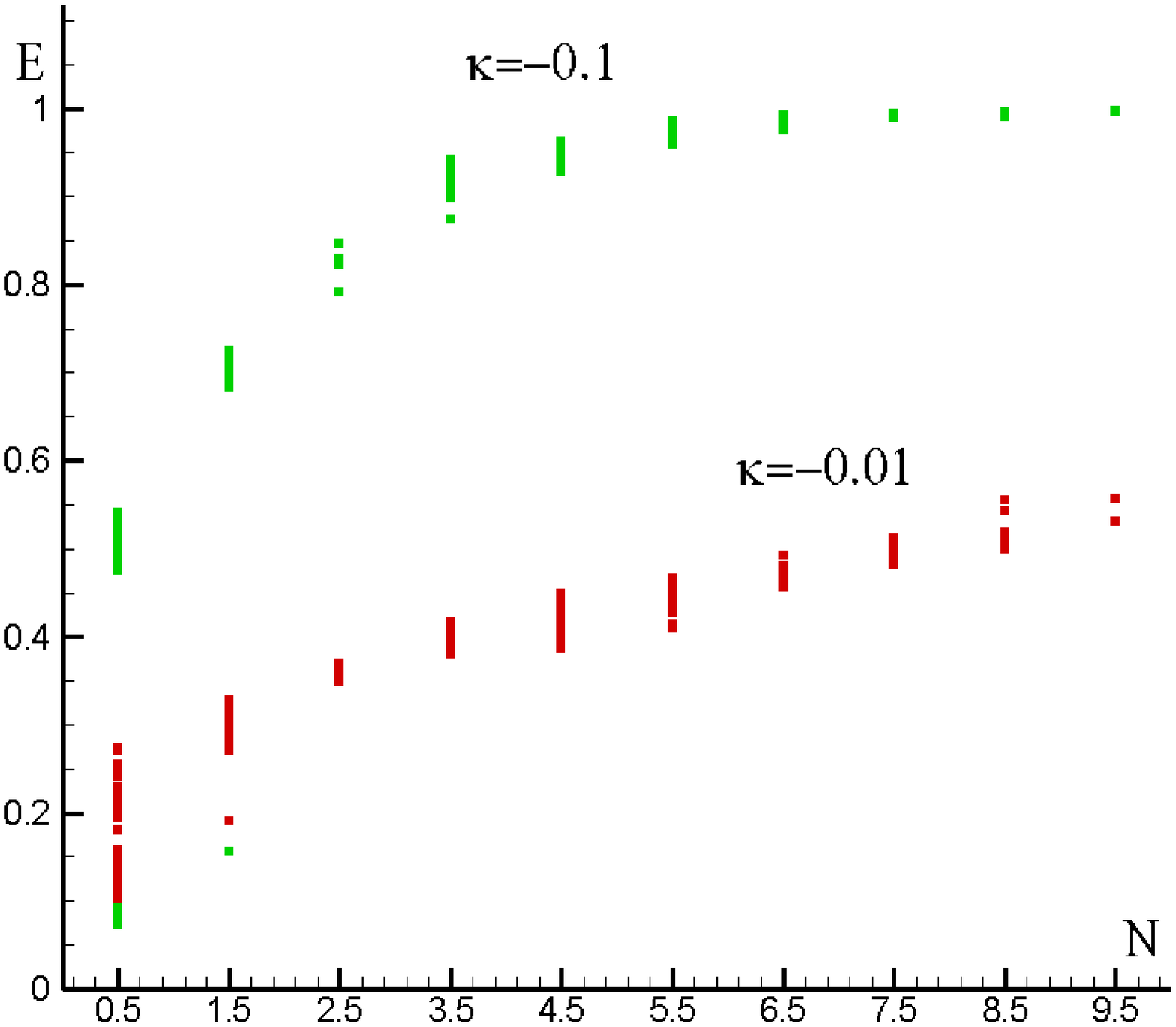}
\caption{upper-layer negative $\kappa$ monopole} 
\end{subfigure}
\begin{subfigure}[t]{0.4\textwidth}
\includegraphics[width=5.3cm]{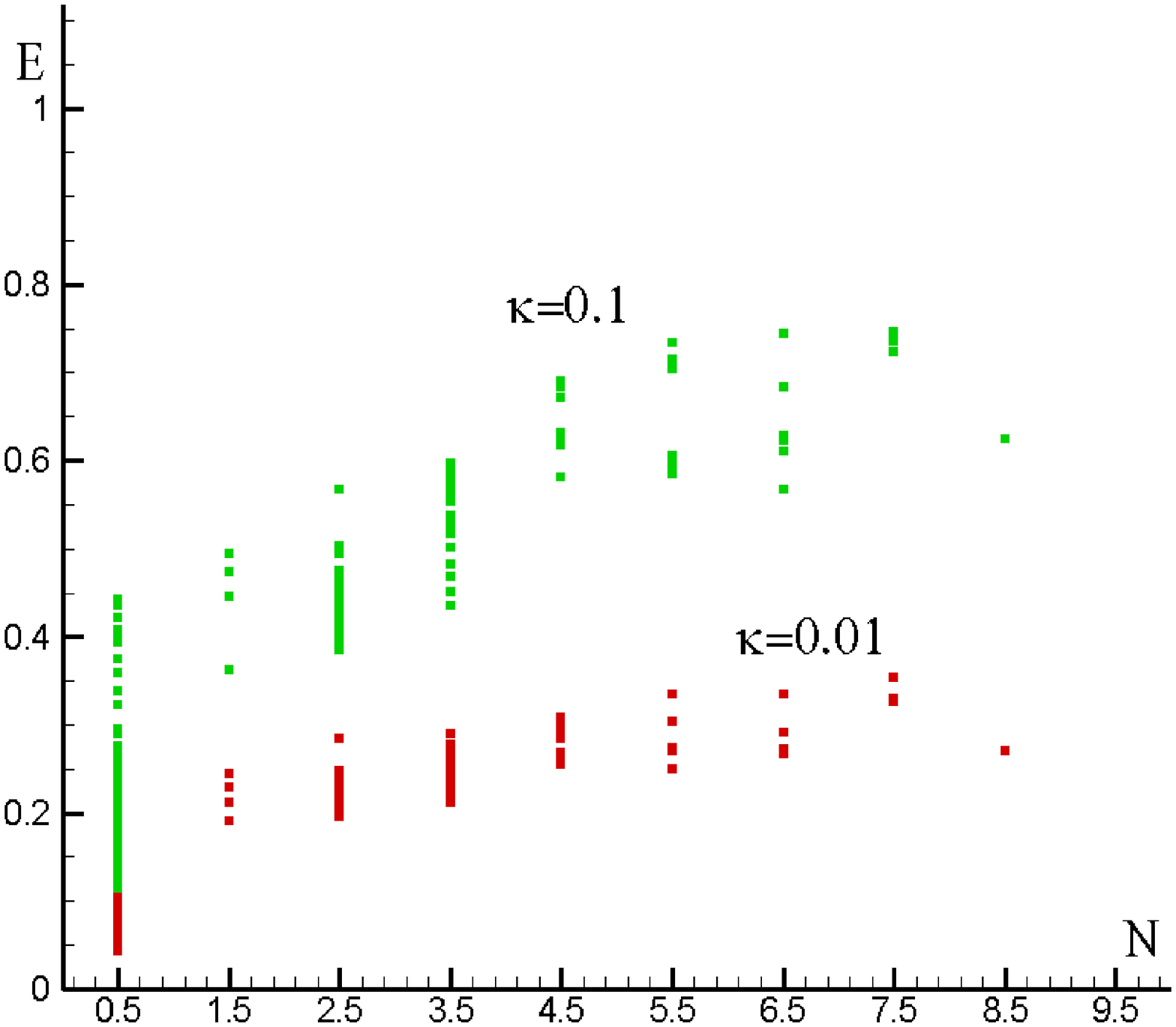}
\caption{middle-layer positive $\kappa$ monopole} 
\end{subfigure}
\quad
\begin{subfigure}[t]{0.4\textwidth}
\includegraphics[width=5.3cm]{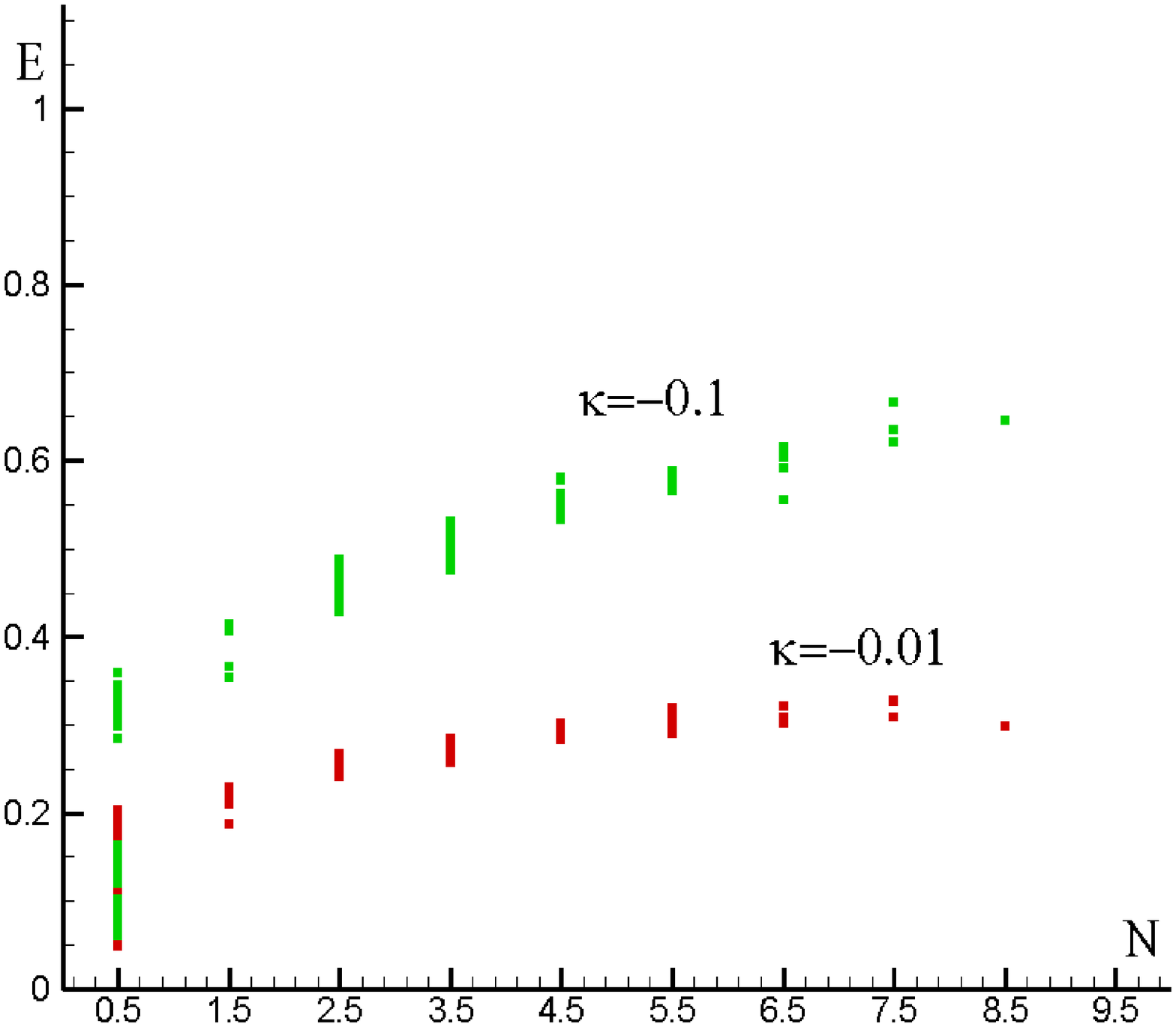}
\caption{middle-layer negative $\kappa$ monopole} 
\end{subfigure}
\end{center}
\caption{Lagrangian advection efficiency $E$ depending on monopole revolution number $N$.}
\label{fig9}
\end{figure*}

Then we have followed the evolution of all the monopoles starting at these initial position, calculating revolution number $N$ of each of those monopoles, and obtained the advection efficiency through expression $E= n_a/n_i$, where $n_a$ is the number of advected out of the topographic vortex markers, i.e. the markers that have crossed line $x=5$, and $n_i=10^4$ is the initial marker distribution number. It is also worth noting, that although some of these monopoles have revolved about the topography equal times, Lagrangian advection being generated by these monopoles is mostly equivalent in each case (see fig. \ref{fig9}). Indeed, each point in fig. \ref{fig9} corresponds to one initial position of the monopole. Thus, if different initial positions correspond to equal number of monopole revolutions $N$, then advection efficiency $E$ is sufficiently similar.

Figure \ref{fig9} depicts advection efficiency $E$ in the upper-layer monopole propagation case (see fig. \ref{fig9}a,b), and in the middle-layer monopole propagation case (see fig. \ref{fig9}c,d). By analyzing these subfigures, one can draw several conclusions. First, $N=0.5$ corresponds to the case of monopole passing very close to the topographic vortex but not being captured by it. In this case, although, if monopole is very weak ($\kappa=0.01$), it causes a great deal of fluid particle advection. A few monopole revolutions are enough for all the particles from the topographic vortex region to be carried away. Second, the sign of the monopole self-rotation is not the main reason of the advection efficiency, but this efficiency is mostly determined by $|\kappa|)$. Third, evidently, a singular monopole (see fig. \ref{fig9}a,b) causes much more efficient advection than a regular one (see fig. \ref{fig9}c,d). 

\conclusions  
In the frame of a three-layer geophysical flow model, Lagrangian advection of fluid particle in the vicinity of a monopole vortex interacting with a topographic vortex has been addressed. Two cases of the monopole propagation have been investigated: the upper-layer monopole propagation, and middle-layer monopole propagation. Such advection has been shown to be determined by two most significant processes. First, chaotic advection due to the nonstationarity of the monopole-topography interaction, and, second, the appearance or disappearance of closed recirculation zones in time. Cooperative influence of these processes causes very effective Lagrangian advection. Two controlling parameters, namely, the monopole's strength and initial position have been analyzed, and, on the basis of the number of regular critical points assessment, a classification of different regimes of Lagrangian advection has been presented. 

By adding a nonstationary term to the background flow, we have analyzed a short-term monopole-topography interaction. If the monopole passes nearly the topographic vortex, it still causes a great deal of particles initially located within the topographic vortex to be carried away. If the monopole is captured by the topographic vortex, then it rotates certain times about the topography, and, finally, is carried away by the background flow. During this passage, the topographic vortex almost completely renews its fluid. 

\begin{acknowledgements}
The reported study was partially supported by RFBR, research projects Nos.: 11-05-00025-A, 11-01-12057-OFI-M, 12-05-
31011.
\end{acknowledgements}

%
%
%
%

\bibliographystyle{copernicus}
\bibliography{monopole_topography_interaction}

\end{document}